\UseRawInputEncoding
\documentclass[twocolumn]{aastex62}
\usepackage{graphicx}
\usepackage{amsmath}

\newcommand{\kms}{{km~s$^{-1}$}}

\newcommand{\lsun}{L$_{\odot}$}
\newcommand{\msun}{M$_{\odot}$}

\graphicspath{{./}{figures/}}

\received{XXX}
\revised{YYY}
\accepted{ZZZ}
\submitjournal{ApJ}

\shorttitle{SN~2013ai}
\shortauthors{Davis et al.}

\begin{document}

\title{SN~2013ai: a link between hydrogen-rich and hydrogen-poor core-collapse supernovae}

\author[0000-0002-2806-5821]{S. Davis}
\affil{Department of Physics, University of California, 1 Shields Avenue, Davis, CA 95616-5270, USA}

\author[0000-0002-8041-8559]{P.J. Pessi}
\affil{Facultad de Ciencias Astron\'{o}micas y Geof\'{i}sicas (FCAG), Universidad Nacional de La Plata (UNLP), Paseo del bosque S/N, 1900, Argentina}
\affil{European Southern Observatory, Alonso de C\'ordova 3107, Casilla 19, Santiago, Chile}

\author{M. Fraser}
\affil{School of Physics, O’Brien Centre for Science North, University College Dublin, Belfield, Dublin 4, Ireland}

\author{K. Ertini}
\affil{Facultad de Ciencias Astron\'{o}micas y Geof\'{i}sicas (FCAG), Universidad Nacional de La Plata (UNLP), Paseo del bosque S/N, 1900, Argentina}
\affil{Instituto de Astrof\'{i}sica de La Plata (IALP), CONICET, Paseo del bosque S/N, 1900, Argentina}

\author[0000-0003-0766-2798]{L. Martinez}
\affil{Facultad de Ciencias Astron\'{o}micas y Geof\'{i}sicas (FCAG), Universidad Nacional de La Plata (UNLP), Paseo del bosque S/N, 1900, Argentina}
\affil{Instituto de Astrof\'{i}sica de La Plata (IALP), CONICET, Paseo del bosque S/N, 1900, Argentina}

\author[0000-0002-4338-6586]{P. Hoeflich}
\affil{Department of Physics, Florida State University, 77 Chieftan Way, Tallahassee, FL 32306, USA}

\author[0000-0003-1039-2928]{E.Y. Hsiao}
\affil{Department of Physics, Florida State University, 77 Chieftan Way, Tallahassee, FL 32306, USA}

\author{G. Folatelli}
\affil{Facultad de Ciencias Astron\'{o}micas y Geof\'{i}sicas (FCAG), Universidad Nacional de La Plata (UNLP), Paseo del bosque S/N, 1900, Argentina}
\affil{Instituto de Astrof\'{i}sica de La Plata (IALP), CONICET, Paseo del bosque S/N, 1900, Argentina}
\affil{Kavli Institute for the Physics and Mathematics of the Universe, Todai Institutes for Advanced Study, University of Tokyo, 5-1-5 Kashiwanoha, Kashiwa, Chiba 277-8583, Japan}

\author[0000-0002-5221-7557]{C. Ashall}
\affiliation{Institute for Astronomy, University of Hawai’i at Manoa, 2680 Woodlawn Dr. Hawai’i, HI 96822, USA}

\author[0000-0003-2734-0796]{M. M. Phillips}
\affiliation{Carnegie Observatories, Las Campanas Observatory, Casilla 601, La Serena, Chile}

\author[0000-0003-0227-3451]{J. P. Anderson}
\affil{European Southern Observatory, Alonso de C\'ordova 3107, Casilla 19, Santiago, Chile}

\author{M. Bersten}
\affil{Facultad de Ciencias Astron\'{o}micas y Geof\'{i}sicas (FCAG), Universidad Nacional de La Plata (UNLP), Paseo del bosque S/N, 1900, Argentina}
\affil{Instituto de Astrof\'{i}sica de La Plata (IALP), CONICET, Paseo del bosque S/N, 1900, Argentina}
\affil{Kavli Institute for the Physics and Mathematics of the Universe, Todai Institutes for Advanced Study, University of Tokyo, 5-1-5 Kashiwanoha, Kashiwa, Chiba 277-8583, Japan}

\author{B. Englert}
\affil{Facultad de Ciencias Astron\'{o}micas y Geof\'{i}sicas (FCAG), Universidad Nacional de La Plata (UNLP), Paseo del bosque S/N, 1900, Argentina}
\affil{Agencia Nacional de Promoción Científica y Tecnológica (ANPCyT), Godoy Cruz 2370, C142FQD, Buenos Aires, Argentina}

\author{A. Fisher}
\affil{Department of Physics, Florida State University, 77 Chieftan Way, Tallahassee, FL 32306, USA}

\author{S. Benetti}
\affil{1INAF - Osservatorio Astronomico di Padova, Vicolo dell’Osservatorio 5, I-35122 Padova, Italy}

\author{A. Bunzel}
\affil{Facultad de Ciencias Astron\'{o}micas y Geof\'{i}sicas (FCAG), Universidad Nacional de La Plata (UNLP), Paseo del bosque S/N, 1900, Argentina}
\affil{Instituto Argentino de Radioastronom\'ia, CC5, 1894 Villa Elisa, Buenos Aires, Argentina}

\author[0000-0003-4625-6629]{C. Burns}
\affiliation{Observatories of the Carnegie Institution for Science, 813 Santa Barbara St., Pasadena, CA 91101, USA }

\author{T.~W. Chen}
\affiliation{The Oskar Klein Centre, Department of Astronomy, Stockholm University, AlbaNova, SE-10691 Stockholm, Sweden}

\author[0000-0001-6293-9062]{C. Contreras}
\affiliation{Carnegie Observatories, Las Campanas Observatory, Casilla 601, La Serena, Chile}

\author{N.~Elias-Rosa}
\affil{INAF Osservatorio Astronomico di Padova, Vicolo dell'Osservatorio 5, 35122 Padova, Italy}
\affil{Institute of Space Sciences (ICE, CSIC), Campus UAB, Carrer de Can Magrans s/n, 08193 Barcelona, Spain}

\author{E.~Falco}
\affiliation{Harvard-Smithsonian Center for Astrophysics, Cambridge, MA 02138, USA}

\author[0000-0002-1296-6887]{L. Galbany}
\affiliation{Departamento de F\'isica Te\'orica y del Cosmos, Universidad de Granada, E-18071 Granada, Spain}

\author[0000-0002-1966-3942]{R. P. Kirshner}
\affiliation{Gordon and Betty Moore Foundation, 1661 Page Mill Road, Palo Alto, CA 94304 }
\affiliation{Harvard-Smithsonian Center for Astrophysics, 60 Garden Street, Cambridge, MA 02138}

\author[0000-0001-8367-7591]{S. Kumar}
\affil{Department of Physics, Florida State University, 77 Chieftan Way, Tallahassee, FL 32306, USA}

\author[0000-0002-3900-1452]{J. Lu}
\affil{Department of Physics, Florida State University, 77 Chieftan Way, Tallahassee, FL 32306, USA}

\author{J.~D. Lyman}
\affil{Department of Physics, University of Warwick, Coventry CV4 7AL, UK}

\author{G.~H.~Marion}
\affiliation{University of Texas at Austin, 1 University Station C1400, Austin, TX, 78712-0259, USA}

\author[0000-0001-7497-2994]{S. Mattila}
\affiliation{Tuorla Observatory, Department of Physics and Astronomy, University of Turku, FI-20014 Turku, Finland}

\author{J. Maund}
\affiliation{Department of Physics and Astronomy, University of Sheffield, F39 Hicks Building, Hounsfield Road, Sheffield, S3 7RH, United Kingdom}

\author[0000-0003-2535-3091]{N. Morrell}
\affiliation{Carnegie Observatories, Las Campanas Observatory, Casilla 601, La Serena, Chile}

\author{J. Ser\'{o}n}
\affiliation{Cerro Tololo Inter-American Observatory/NSF’s NOIRLab, Casilla 603, La Serena, Chile}

\author[0000-0002-5571-1833]{M. Stritzinger }
\affiliation{Department of Physics and Astronomy, Aarhus University, 
Ny Munkegade 120, DK-8000 Aarhus C, Denmark }

\author[0000-0002-9301-5302]{M. Shahbandeh}
\affiliation{Department of Physics, Florida State University, 77 Chieftan Way, Tallahassee, FL 32306, USA}

\author{M. Sullivan}
\affil{School of Physics and Astronomy, University of Southampton, Southampton, SO17 1BJ, UK}

\author[0000-0002-8102-181X]{N. B. Suntzeff}
\affiliation{George P. and Cynthia Woods Mitchell Institute for Fundamental Physics \& Astronomy, Texas A\&M University, Department of Physics, 4242 TAMU, College Station, TX 77843}

\author{D.~R. Young}
\affil{Astrophysics Research Centre, School of Mathematics and Physics, Queens University Belfast, Belfast BT7 1NN, UK}

\correspondingauthor{Scott Davis}
\email{sfdav@ucdavis.edu}

\begin{abstract}
We present a study of optical and near-infrared (NIR) spectra along with the light curves of SN~2013ai. These data range from discovery until 380 days after explosion.
SN~2013ai is a fast declining type II supernova (SN~II) with an unusually long rise time; $18.9\pm2.7$d in $V$-band and a bright $V$-band peak absolute magnitude of $-18.7\pm0.06$ mag. The spectra are dominated by hydrogen features in the optical and NIR. The spectral features of SN~2013ai are unique in their expansion velocities, which when compared to large samples of SNe~II are more than 1,000 \kms\ faster at 50 days past explosion. In addition, the long rise time  of the light curve more closely resembles SNe~IIb rather than SNe~II. If SN~2013ai is coeval with a nearby compact cluster we infer a progenitor ZAMS mass of $\sim$17 M$_\odot$. After performing light curve modeling we find that SN~2013ai could be the result of the explosion of a star with little hydrogen mass, a large amount of synthesized $^{56}$Ni, 0.3-0.4 \msun, and an explosion energy of $2.5-3.0\times10^{51}$ ergs.
The density structure and expansion velocities of SN~2013ai are similar to that of the prototypical SN~IIb, SN~1993J. However, SN~2013ai shows no strong helium features in the optical, likely due to the presence of a dense core that prevents the majority of $\gamma$-rays from escaping to excite helium. Our analysis suggests that SN~2013ai could be a link between SNe~II and stripped envelope SNe.
\end{abstract}

\section{Introduction}

Core collapse supernovae (CCSNe) are the result of the explosion of massive stars ($>8$ \msun). These explosions are characterized by two main classes: those that show no hydrogen and those that do, type I (SNe~I) and type II supernovae (SNe~II), respectively \citep{Minkowski1941}. 

Type I CCSNe are produced when a massive star is stripped of most of its outer hydrogen and possibly its helium layers before explosion, thus referred to as stripped envelope SNe \citep[SESNe,][]{Clocchiatti1997}. It is believed that the outer layers of the SESNe progenitors are primarily removed either through strong winds \citep{Woosley1993} or binary interaction \citep{Nomoto1995,Podsiadlowksi2004}. These SESNe are further divided into two general spectroscopic groups, SNe~Ib that show strong helium lines in their optical spectra and SNe Ic that do not. SNe~Ib and Ic form spectroscopically homogeneous groups with minimal variation \citep{Modjaz2014}. The light curves of SNe~Ib and Ic are similar and can be characterized by a slow $\sim20$ day rise to maximum followed by a post-maximum decline of $\sim0.75$ magnitudes per 15 days in the $V$-band \citep[e.g.][]{Drout2011,Taddia2018,Prentice2019}.


When a massive star retains most of its hydrogen envelope, it explodes as a SN~II. SNe~II were historically divided into groups based on the shape of their light curves. Those with a slow decline -- a plateau -- after maximum were classified as IIP and those with no plateau after maximum were classified as IIL \citep{Barbon1979}. However, recent studies have shown that there is likely a continuum in the decline rates and suggest that slow (IIP) and fast decliners (IIL) are not distinct groups \citep{Anderson2014,Sanders2015, Galbany2016,Valenti2016,Rubin2016,Pessi2019,deJaeger2019}. The different decline rates were proposed to correlate with the amount of hydrogen retained by the progenitor \citep[e.g.][]{Popov1993,Anderson2014,Faran2014,Gutierrez2014,Moriya2016,Hillier2019}; however, see \citep{Morozova2017} for an alternative explanation based on interaction with dense circumstellar material. On the contrary,  the near-infrared (NIR) shows that there are spectroscopic distinctions between slow and fast decliners in the 1.06$\mu m$ region, which may be due to progenitor differences \citep{Davis2019}.

Pre-explosion images of SNe~IIP suggest that most of their progenitors are red supergiants (RSG) \citep[e.g.][]{Smartt2004,Maund2005,Smartt2009,Smartt2015,Vandyk2017,Vandyk2019}. However, blue supergiants can produce SNe~II with longer rising light curves called SN~1987A-like after the famous SN~1987A (see \citealt{Arnett1989} for a review). Theory suggests that if a RSG progenitor has most of its hydrogen envelope stripped pre-explosion, it can give rise to a SN~IIb; however, the progenitor likely plays a role as it may not be that of a normal SN~II (see \citealt{Bersten2012}). SN~IIb show hydrogen at early times and helium in their later spectra that quickly becomes the dominant feature \citep{Filippenko1993}. The required strong mass loss is believed to be due to a binary progenitor \citep[e.g.][]{Podsiadlowski1993,Aldering1994}. 

There is an open question as to whether or not there is a continuum between fast declining SN~II and SN~IIb. \citet{Arcavi2012} and \citet{Pessi2019} suggest that there is a discrete change between SN~II and SN~IIb. However, \citet{Pessi2019} discussed a peculiar SN~II, SN 2013ai, that presented an unusually large rise time and was systematically ``misclasified" as a SN~IIb when performing clustering analysis. This, along with the spectral peculiarities discussed later in the current work, encouraged further analysis of SN~2013ai in order to elucidate if the object is in fact some kind of transitional event, in which case, it can provide important clues about the stellar evolutionary pathways leading to CCSNe.


In this paper we present optical and NIR observations and analysis of  SN~2013ai. Section \ref{obs} outlines the observing techniques and reduction procedures. In Section \ref{reddening} we describe the processes used to determine the reddening of SN~2013ai. In Sections \ref{sec:phot} and \ref{spec} we present the photometric and spectroscopic properties of SN~2013ai and make comparisons to other CCSNe. The analysis of the pre-explosion HST images and the possible progenitor scenarios are described in Section \ref{sect:progenitor}. In Section \ref{sect:model} the models and their outcomes are reviewed. The discussion of results and conclusions are in Sections \ref{sect:discussion} and \ref{sect:conclusions}, respectively.

\section{Observations and Data Reduction}
\label{obs}

SN~2013ai was discovered at RA(2000)$=6^h16^m18^s.35$, Dec(2000)$=-21^{\circ}22^\prime32^{\prime \prime}.90$ with the Zadko 1 meter telescope at Gingin Observatory, Australia on 2013 March 1 UT at an apparent $R$-band magnitude of 14.4 mag \citep{atel4849}. However, the SN was previously detected in pre-discovery images as early as 2013 February 26 UT and not detected in images taken on 2013 February 21 UT, with a limiting magnitude of $r=20.7$ from the Panoramic Survey Telescope and Rapid Response System \citep[Pan-STARRS;][]{Chambers2016}. Thus an explosion date of 2013 February 24 UT (56347 MJD) $\pm2.7$ days is adopted for the remainder of this work. The error listed on explosion date is a flat distribution without any obvious bias toward last non-detection or discovery. 

The host of SN~2013ai, NGC 2207, is in the process of colliding with another galaxy, IC 2163, which makes any distance measurement based on galaxian properties uncertain. The host is a luminous infrared galaxy with a high star formation rate that produces SNe frequently, e.g. SNe~1975A, 1999ec, 2003H, 2018lab, and AT~2019eez \citep{Kirshner1976,Jha1999,Graham2003,Sand2018,Stanek2019}.
At a redshift of 0.009, NGC 2207 is too close for a reliable redshift distance as it is not in the Hubble flow. 


NGC 2207 was host to SN 1975A, a SN Ia that provides the only redshift-independent distance measurement to the galaxy. Using the GELATO classification tool \citep{Harutyunyan2008}, the best match of the 1975 January 20 spectrum of SN 1975A published by \citet{Kirshner1976} was found to be SN2003du, which had a $B$-band decline rate of $\Delta{\rm{m}_{15}}(B)=1.07\pm0.06$ \citep{Blondin2012}. Pseudo equivalent width (pEW) measurements of the Si~{\sc ii} $\lambda\lambda$5972 and 6355 lines from this spectrum place SN 1975A in the middle of core-normal SNe in the Branch diagram \citep{Branch2006,Branch2009}. In addition, the pEW of the λ5972 line is consistent with $\Delta{\rm{m}_{15}}(B)\sim$1.1 (see Figure 17b of \citealt{Folatelli2013}). Following the procedures of Riess et al. (1998), and using the reproduced photometry from \citet{Kirshner1976}, the SN 1975A light curves were analyzed using SNooPy \citep{Burns2011} to estimate a distance. Fixing the B-band decline rate in the range $\Delta{\rm{m}_{15}}$(B)$=1.1-1.5$ yielded a 0.1 mag difference in the distance moduli and a fit error of 0.15 mag from SNooPy.  This procedure gives a final distance modulus of $33.5 \pm 0.2$ for NGC 2207, which we adopt in this paper.

SN~2013ai was classified as a SN~II by the Public ESO Spectroscopic Survey of Transient Objects (PESSTO) \citep{Smartt2015} due to broad H$\alpha$ emission in its optical spectrum \citep{cbet3431}. Archival HST-WFPC2 images are available of the field, obtained using the F336W, F439W, F555W, and F814W filters, which contained the site of SN~2013ai on 1996 May 25 UT. However, no clear single progenitor could be detected \citep{atel4862}. These data are reanalyzed in Section \ref{sect:progenitor}.

\begin{figure}
    \includegraphics[width=\columnwidth]{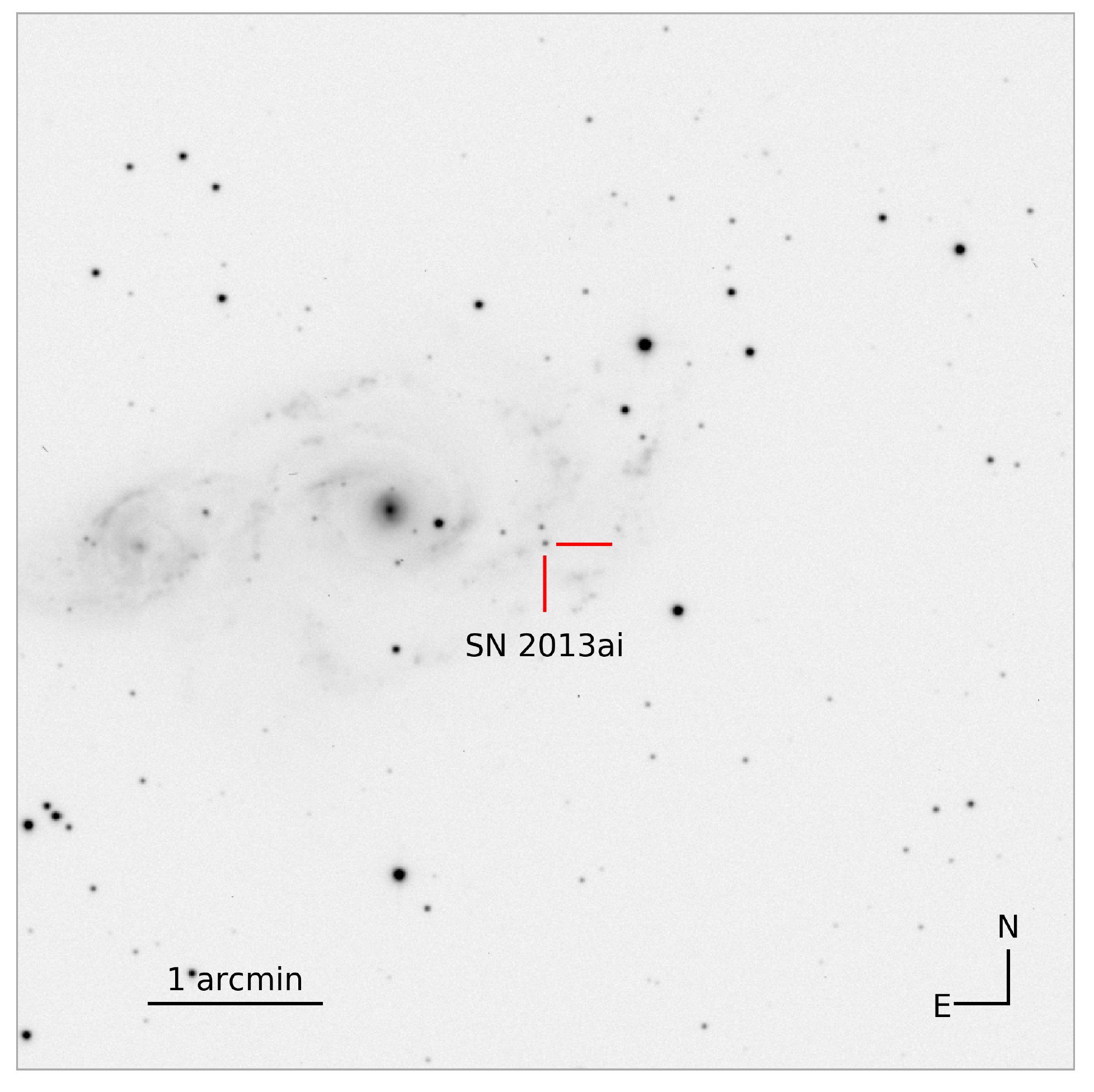}
    \caption{Swope $V$-band image of SN~2013ai in NGC 2207 taken around 10 days past explosion. The cross hair marks the SN. The compass and size of field are also noted.}
    \label{fig:hostImage}
\end{figure}

Optical photometric follow up commenced soon after SN~2013ai was discovered, using ESO + EFOSC2, the 1.3~m SMARTS telescope\footnote{Operated by the SMARTS Consortium.} + ANDICAM, and the 1m Swope telescope + SITe3. 
Swope + SITe3 images were taken as a part of the Carnegie Supernova Project II (CSP-II) \citep{Phillips2019,Hsiao2019}.
In the NIR, imaging was obtained with NTT + SOFI and the 2.5~m du Pont telescope + RetroCam. A Swope $V$-band image taken at $\sim10$ days past explosion is shown in Figure \ref{fig:hostImage}.

ANDICAM data were reduced with the dedicated pipeline which subtracts the overscan and bias from each image, followed by flat fielding correction using dome flats. For EFOSC2 data, the reduction was performed using the PESSTO pipeline, as described in \citet{Smartt2015}, which trims, debiases, and flat-fields images (using twilight sky flats).  In addition, a fringe frame is used to correct the {\it I}-band data. The SN magnitude in each frame was measured via Point Spread Function (PSF) fitting photometry using the {\sc SNOoPy} package\footnote{SNOoPy is a package for SN photometry using PSF fitting and/or template subtraction developed by E. Cappellaro. A package description can be found at https://sngroup.oapd.inaf.it/snoopy.html}, with errors estimated from artificial star tests. The photometric zero point for each image was determined from aperture photometry of local sequence stars, which in turn were calibrated to Landolt fields \citep{Landolt1992} observed on multiple photometric nights. Color terms were applied to the EFOCS2 and ANDICAM magnitudes using the values listed in \citet{Smartt2013} for the former and values listed on the SMARTS consortium webpages for the latter. 
The reduction of Swope photometry was performed as described in \citet{Phillips2019}. Swope PSF photometry was performed in the same process described for ANDICAM and EFOSC2 data with standard fields observed on the same nights as the SN observations when possible. The Swope data have been converted to the standard system using the process outlined in \citet{Phillips2019}.
The light curves are presented without template subtraction as the SN is well separated from the host. A log of the optical photometry is given in Table \ref{tab:optPhot}.

\begin{deluxetable*}{llcccccccr}
\tablenum{1}
\tablecaption{Optical Photometry of SN~2013ai.}
\centering
\label{tab:optPhot}
\tablecolumns{10}
\tablehead{
\colhead{MJD} &
\colhead{UT Date} &
\colhead{Phase} &
\colhead{$u$ ($\sigma$)} &
\colhead{$B$  ($\sigma$)} &
\colhead{$V$ ($\sigma$)} &
\colhead{$g$ ($\sigma$)} &
\colhead{$R$ ($\sigma$)} &
\colhead{$I$ ($\sigma$)} &
\colhead{Instrument}
}
\startdata
56344.33 & 2013/02/21 & $-$2.7 & - & - & - & - & $>$20.7 & - & GPC1 \\
56349.66 & 2013/02/26 & 2.7 & - & - & - & - & 18.3 & - & TAROT \\
56351.66 & 2013/02/28 & 4.7 & - & - & - & - & 17.6 & - & TAROT \\
56352.66 & 2013/03/01 & 5.7 & - & - & - & - & 17.4 & - & TAROT \\
56354.05 & 2013/03/03 & 7.1 & - & - & - & - & 17.23(03) & - & EFOSC2\\
56354.25 & 2013/03/03 & 7.2 & 18.43(04) & 18.15(02) & 17.68(02) & 17.80(02) & 17.42(02) & 17.17(02) & SITe3\\
56355.08 & 2013/03/04 & 8.1 & 18.22(03) & 18.07(02) & 17.58(02) & 17.80(02) & 17.27(03) & 17.06(02) & SITe3\\
56356.09 & 2013/03/05 & 9.1 & - & 18.22(02) & 17.49(02) & - & 16.94(03) & 16.35(04) & ANDICAM\\
56357.08 & 2013/03/06 & 10.1 & - & 18.09(01) & 17.45(02) & - & 16.83(01) & 16.21(02) & ANDICAM\\
56362.15 & 2013/03/11 & 15.2 & - & 18.09(03) & 17.30(02) & - & 16.69(03) & 16.06(03) & ANDICAM\\
56362.15 & 2013/03/11 & 15.2 & - & - & 17.34(04) & - & - & - & EFOSC2\\
56364.15 & 2013/03/13 & 17.2 & - & 18.14(03) & 17.29(02) & - & 16.70(03) & 16.05(03) & ANDICAM\\
56364.15 & 2013/03/13 & 17.2 & - & - & 17.36(03) & - & - & - & EFOSC2\\
56368.12 & 2013/03/17 & 21.1 & - & 18.21(02) & 17.36(02) & - & 16.65(02) & 15.98(02) & ANDICAM\\
56368.08 & 2013/03/17 & 21.1 & - & - & 17.37(03) & - & - & - & EFOSC2\\
56371.13 & 2013/03/20 & 24.1 & - & 18.33(03) & 17.34(04) & - & 16.60(01) & 15.93(04) & ANDICAM\\
56373.14 & 2013/03/22 & 26.1 & - & 18.36(07) & 17.40(04) & - & 16.59(02) & 15.86(02) & ANDICAM\\
56374.10 & 2013/03/23 & 27.1 & - & 18.46(04) & 17.40(04) & - & 16.54(04) & 15.96(03) & ANDICAM\\
56378.05 & 2013/03/27 & 31.1 & - & 18.73(06) & 17.47(04) & - & 16.64(02) & 15.95(03) & ANDICAM\\
56380.12 & 2013/03/29 & 33.1 & - & 18.73(08) & 17.54(04) & - & 16.69(03) & 16.02(03) & ANDICAM\\
56385.03 & 2013/04/03 & 38.0 & - & - & 17.75(10) & - & - & - & EFOSC2\\
56388.07 & 2013/04/06 & 41.1 & - & 19.12(04) & 17.68(03) & - & 16.73(04) & 16.07(02) & ANDICAM\\
56391.07 & 2013/04/09 & 44.1 & - & 19.25(05) & 17.71(03) & - & 16.79(02) & 16.02(02) & ANDICAM\\
56394.06 & 2013/04/12 & 47.1 & - & 19.20(42) & 17.81(06) & - & 16.82(04) & 16.06(03) & ANDICAM\\
56395.00 & 2013/04/13 & 48.0 & - & - & 17.91(03) & - & - & - & EFOSC2\\
56397.05 & 2013/04/15 & 50.1 & - & 19.54(08) & 17.80(03) & - & 16.82(03) & 16.17(03) & ANDICAM\\
56400.06 & 2013/04/18 & 53.1 & - & 19.47(36) & 17.92(05) & - & 16.87(06) & 16.09(04) & ANDICAM\\
56402.02 & 2013/04/20 & 55.0 & - & - & 18.00(04) & - & - & - & EFOSC2\\
56403.03 & 2013/04/21 & 56.0 & - & 19.57(09) & 17.98(05) & - & 17.02(03) & 16.19(02) & ANDICAM\\
56411.08 & 2013/04/29 & 64.1 & - & 19.60(60) & - & - & 17.10(03) & - & EFOSC2\\
56414.51 & 2013/05/02 & 67.5 & - & 19.66(62) & 18.16(17) & - & 17.08(03) & 16.37(03) & ANDICAM\\
56680.17 & 2014/01/23 & 333.2 & - & - & - & - & - & 21.57(22) & EFOSC2\\
56695.16 & 2014/02/07 & 348.2 & - & - & 23.75(63) & - & - & 22.33(41) & EFOSC2\\
56708.15 & 2014/02/20 & 361.2 & - & - & 23.21(31) & - & - & - & EFOSC2\\
56726.09 & 2014/03/10 & 379.1 & - & - & - & - & - & 21.48(28) & EFOSC2\\
\enddata
\tablecomments{The MJD column lists the Modified Julian Date of each observation. Phase is listed in days since explosion.}
\end{deluxetable*}

\begin{figure}
    \includegraphics[width=\columnwidth]{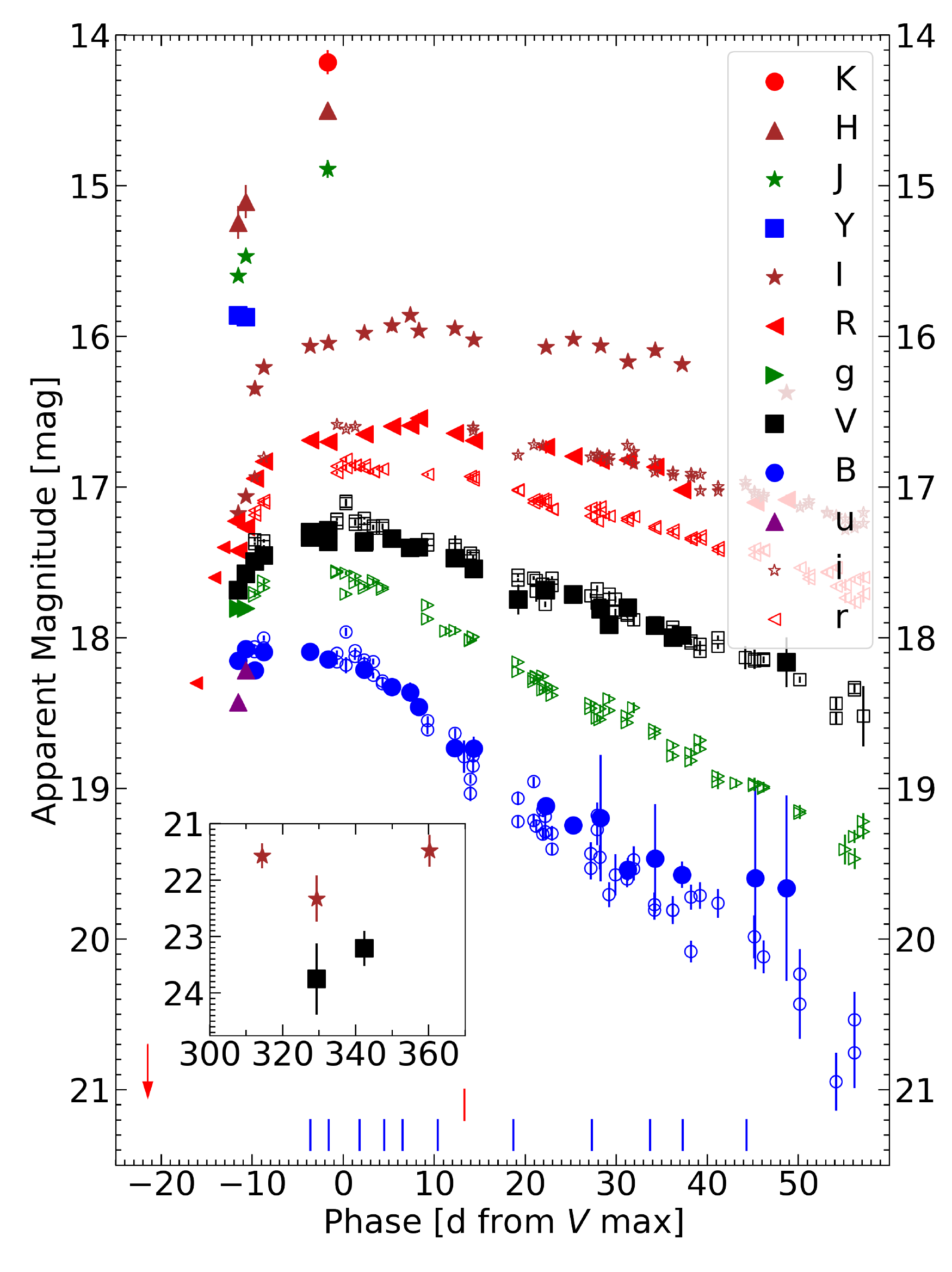}
    \caption{Milky Way corrected optical and NIR light curves of SN~2013ai. Light curves are presented without host subtractions. Blue and red vertical lines mark the dates when optical and NIR spectra were taken, respectively. Open symbols are previously published data from \citet{Valenti2016}. The arrow denotes the magnitude limit of the last non-detection. The inset shows data taken more than 300 days from $V$-band maximum.}
    \label{fig:phot}
\end{figure}

The NIR light curves from Dupont + RetroCam were reduced in the standard manner following \citet{Phillips2019}. The NIR NTT + SOFI data were corrected for crosstalk, flat-fielded (using dome flats), and scattered light, performed within the PESSTO pipeline. Separate off-target sky frames were taken as part of the sequence of observations, and these were used to subtract the sky background from the on-target images. The multiple exposures taken in each filter were combined to produce a deep image on which photometry was performed. The log of NIR photometry is given in Table \ref{tab:nirPhot}. The photometry of SN~2013ai previously published by \citet{Valenti2016} is also included. The light curves in each band, during the photospheric phase, are presented in Figure \ref{fig:phot}.

\begin{deluxetable*}{llcccccr}
\tablenum{2}
\tablecaption{NIR Photometry of SN~2013ai.}
\centering
\label{tab:nirPhot}
\tablecolumns{8}
\tablehead{
\colhead{MJD} &
\colhead{UT Date} &
\colhead{Phase} &
\colhead{$Y$  ($\sigma$)} &
\colhead{$J$ ($\sigma$)} &
\colhead{$H$ ($\sigma$)} &
\colhead{$K_S$ ($\sigma$)} &
\colhead{Instrument}
}
\startdata
56354.25 & 2013/03/03 & 7.1 & 15.86(01) & 15.60(01) & 15.24(11) & - & RetroCam \\
56355.08 & 2013/03/04 & 8.1 & 15.87(01) & 15.47(01) & 15.11(11) & - & RetroCam \\
56364.10 & 2013/03/13 & 17.1 & - & 14.89(06) & 14.50(03) & 14.18(08) & SOFI \\
\enddata
\tablecomments{The MJD column lists the Modified Julian Date of each observation. Phase is given in days since explosion.}
\end{deluxetable*}

\begin{figure}
    \includegraphics[width=\columnwidth]{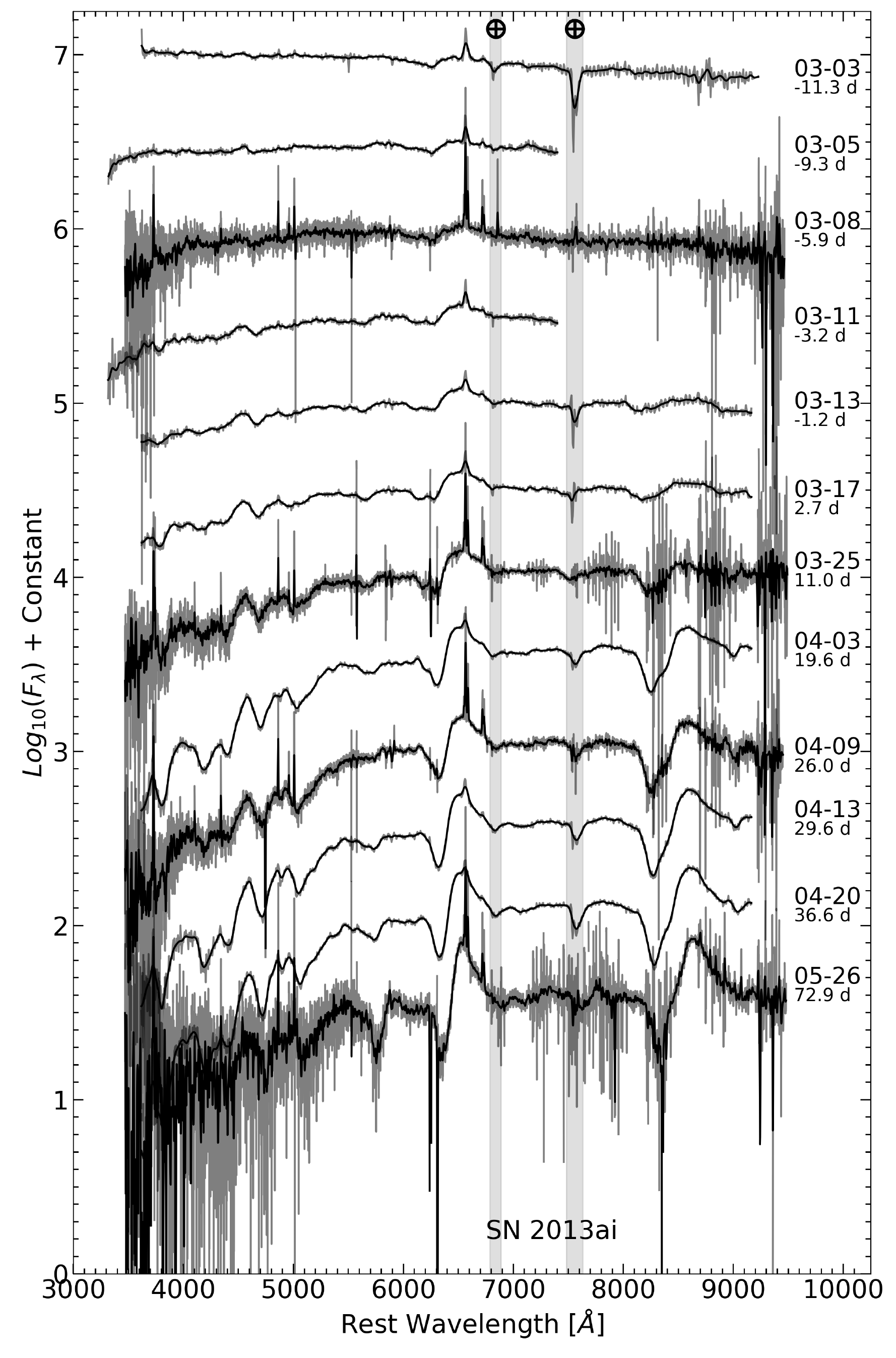}
    \caption{Optical spectra of SN~2013ai. Gaussian smoothed spectra, with a radius of 30 \AA, are plotted (black) over the observed spectra (grey). The UT date of observation and phase relative to $V$ maximum are labeled for each spectrum. The gray vertical bands mark the regions of strongest telluric absorption. The narrow peaks in the H$\alpha$ profile are from the host galaxy of SN~2013ai.}
    \label{fig:optSpec}
\end{figure}

Through the PESSTO collaboration, an optical spectroscopic time-series was obtained with EFOSC2 on the ESO-NTT \citep{Buzzoni1984}. Complementary, spectra were taken with the Wide-Field Spectrograph \citep[WiFeS;][]{Dopita2010} on the Australian National University (ANU) 2.3m Telescope as part of the ANU WiFeS Supernova Programme \citep{Childress2016}. The EFOSC2 spectra were reduced, extracted, and calibrated with the PESSTO pipeline using standard techniques \citep{Smartt2015}. All spectra were bias-subtracted and divided by a normalized lamp flat. For the Gr\#16 EFOSC2 spectra, flat fields were taken immediately after the spectrum, and at the same position on the sky, to allow for the removal of the strong fringing seen at longer wavelengths. Wavelength calibration was performed with respect to an HeAr arc lamp. After extracting each spectrum the wavelength of the strong night sky emission lines were checked, and if necessary, a linear shift was applied to the dispersion solution. 
The spectra were flux calibrated using a sensitivity curve constructed from observations of spectrophotometric standard stars, and then telluric absorptions were removed using a scaled synthetic model of atmospheric transmission \citep{Patat2011}.
The WiFeS spectra were reduced using the {\sc PyWiFeS} software \citep{Childress2013}. The late time spectrum taken with Gran Telescopio Canarias (GTC) on October 30 2013 was reduced and extracted using standard {\sc IRAF} \citep{Tody1986} routines and was flux calibrated using a spectrophotometric standard star observed on the same night. A log of the spectroscopic observations is presented in Table \ref{tab:optLog}, and the optical spectra during the photospheric phase are presented in Figure \ref{fig:optSpec}. 

A NIR spectrum of SN~2013ai was obtained using the Folded-port Infrared Echellette \citep[FIRE;][]{Simcoe2013} on the Magellan Baade telescope as part CSP-II \citep{Phillips2019,Hsiao2019}. The spectrum was reduced and telluric corrected following the procedure outlined in \citet{Hsiao2019} and has been previously published in \citet{Davis2019}. All observations will be made public via WISeREP \citep{Yaron2012}.

Further observations taken with Swift-XRT 25 days after explosion detected an X-ray source 8$^{\prime\prime}$ from the SN position at a level of $\sim2\times10^{-14}$~erg/s/cm$^2$ \citep{Margutti2013}. Radio observations were taken using the Combined Array for Research in Millimeter-Wave Astronomy (CARMA) 10 and 11 days past explosion at 85~GHz, however, no radio source was detected at either epoch at a limit of 0.6~mJy \citep{Zauderer2013}.

\begin{deluxetable}{ccccc}
\tablecaption{Journal of spectroscopic observations}
\tablenum{3}
\label{tab:optLog}
\tablecolumns{5}
\tablewidth{0pt}
\tablehead{
\colhead{UT Date} &  
\colhead{MJD} &
\colhead{Instrument} &
\colhead{Phase} &
\colhead{$t_{max}(V)$}
}
\startdata
& & Optical & & \\
\tableline
2013-03-03 & 56354.1 & NTT + EFOSC2 & 7.1 & $-$11.3 \\
2013-03-05 & 56356.1 & NTT + EFOSC2 & 9.1 & $-$9.3 \\
2013-03-08 & 56359.5 & ANU + WiFeS  & 12.0 & $-$5.9 \\
2013-03-11 & 56362.2 & NTT + EFOSC2 & 15.2 & $-$3.2 \\
2013-03-13 & 56364.2 & NTT + EFOSC2 & 17.2 & $-$1.2 \\
2013-03-17 & 56368.1 & NTT + EFOSC2 & 21.1 & 2.7 \\
2013-03-25 & 56376.4 & ANU + WiFeS  & 29.4 & 11.0 \\
2013-04-03 & 56385.0 & NTT + EFOSC2 & 38.0 & 19.6 \\
2013-04-09 & 56391.4 & ANU + WiFeS  & 44.4 & 26.0 \\
2013-04-13 & 56395.0 & NTT + EFOSC2 & 48.0 & 29.6 \\
2013-04-20 & 56402.0 & NTT + EFOSC2 & 55.0 & 36.6 \\
2013-05-26 & 56438.3 & ANU + WiFeS  & 91.3 & 72.9 \\
2013-10-30 & 56595.5 & GTC + OSIRIS & 249.5 & 231.1 \\
\tableline
& & NIR & & \\
\tableline
2013-03-10 & 56361.5 & Baade + FIRE & 14.5 & $-$3.9
\enddata
\tablecomments{The MJD column lists the Modified Julian Date of each observation. Phase is given as days since explosion. Time relative to $V$ maximum, in days, is denoted $t_{max}(V)$.}
\end{deluxetable}

\section{Reddening}
\label{reddening}

Milky Way extinction towards SN~2013ai is taken from \citep{Schlafly2011}. In order to estimate the extinction of SN~2013ai the equivalent width (EW) of the Na~I absorption lines at the redshift of the host galaxy NGC 2207 were measured in all available WiFeS spectra as they have the highest resolution of R$\sim 7000$. Assuming Gaussian profiles for both components, we obtain an average of $D_{1}= 0.68 \pm 0.15$~\AA, $D_{2}=0.91 \pm 0.07$~\AA, and $D_{1}+D_{2}=1.52 \pm 0.07$~\AA. Using equation (9) from \citet{Poznanski2012}  we obtain an $E(B-V)=0.84 \pm 0.64$~mag. However, according to \citet{Phillips2013} the error associated with this method can be much larger than predicted by relations derived for typical Milky Way dust and gas.
As a check for consistency, we used the Balmer decrement to determine the dust extinction by computing the flux ratio of H$\alpha$ and H$\beta$ lines using the same WiFes spectra. The Balmer decrement was measured using the narrow host galaxy lines in the spectrum.
The obtained color excess was $E(B-V)=0.70 \pm 0.34$~mag. Similar results were obtained using PESSTO low resolution spectra from which we extracted a region near the SN without subtracting the background. The Balmer decrement was measured from a flux calibrated MW extinction corrected spectrum. Errors were determined by a series of Monte Carlo realizations, assuming the uncertainty is dominated by the photometric errors and therefore flux calibration. The Balmer decrement was measured for each realization, with the error taken as the standard deviation. Finally, the diffuse interstellar band (DIB) at 5780~\AA\ was analyzed. This band appears to be very weak or non existent in the available spectra.
Thus, for this work we adopt a color excess estimate of $E(B-V)_{host}\sim0.8 \pm 0.5$~mag from the Balmer decrement and Na ID measurements.


\section{Photometric Properties}
\label{sec:phot}

The optical light curves of SN~2013ai have been previously studied by \citet{Valenti2016} and \citet{Pessi2019}. The $V$-band light curve was characterized by \citet{Davis2019} using the parameters defined by \citet{Anderson2014}. 
Here we combine the public data of SN~2013ai along with a previously unpublished data set to further analyze SN~2013ai.

The new data are in agreement with previous results that positioned SN~2013ai among typical fast declining SNe~II with a decline rate per 100 days, $s_{2}$ \citep{Anderson2014}, of $\sim 2.0$ \citep[historically SNe~IIL, see][]{Barbon1979}; and an absolute $V$-band maximum of $-18.7\pm0.06$ mag. The only photometric difference SN~2013ai displays with respect to normal SNe~II is that it presents an atypically long rise time \citep[e.g.][]{Valenti2016}. From the \citet{Pessi2019} sample, the typical rise times of SNe~II are $8.3\pm2.0$ days, $12.8\pm2.4$ days, and $16.0\pm3.6$ days in the $B$, $V$, and $r$ bands, respectively. 
For SN~2013ai, the rise times are 13.0$\pm$5.3 days, 18.9$\pm$3.7 days, and 24.4$\pm$4.0 days, in the  $B$, $V$, and $R$ bands, respectively.
On average the rise times of SN~2013ai are 2.4$\sigma$ away from the mean values calculated in \cite{Pessi2019}. Note that in \citet{Pessi2019} the computed rise time average for the SN~II sample is longer than for other higher cadence samples \citep[e.g.][]{Gonzalez2015,Gall2015,Valenti2016,Rubin2016}.
\citet{Rubin2016} studied the $R$-band light curves of 44 SNe~II and found an average rise time of 7.8$\pm$2.8 days with the slowest rising object, PTF12hsx, taking 16.2 days to reach maximum. SN~2013ai rises significantly more slowly than any SN in the \citet{Rubin2016} sample. The rise time of SN~2013ai is significantly longer than that of a typical SNe~II.



Late-time photometric data may be used to estimate the $^{56}$Ni mass using the procedure of \citet{Hamuy2003} and the bolometric correction from \citet{Bersten2009}. However, the scarce late-time data do not allow for a decay slope to be measured, so only a lower limit of the $^{56}$Ni mass can be given as the extent of the $\gamma$-ray trapping can only be estimated. For 100\% trapping, we get $0.14^{+0.06}_{-0.04}$ \msun\  with the error bars including the photometry error but not the reddening and distance uncertainties. There are large uncertainties associated with this lower limit as the light curve decay is highly dependent on the amount of $\gamma$-ray trapping. For the case of SN~1993J \citep[see][]{Hoeflich1993}, which is a good approximation for SN~2013ai at late times (see Section \ref{sect:model}), 79\% of $\gamma$-rays had escaped at 350 days past explosion. However, $^{56}$Co decays in two channels, electron capture and positron capture. Approximately 4\% of the decays are positron capture and will remain trapped, giving a total trapping of $\sim25$\%. This suggests that the $^{56}$Ni mass lower limit from observations could be underestimated by up to a factor of 4, translating to a $^{56}$Ni mass of $0.4-0.8$ \msun. Similarly, assuming that the light curve is powered by $^{56}$Co decay past 50 days from explosion, i.e. constant slope, the magnitude difference of pure $^{56}$Co decay to the observed light curve gives a similar result.

\begin{figure}
    \includegraphics[width=\columnwidth]{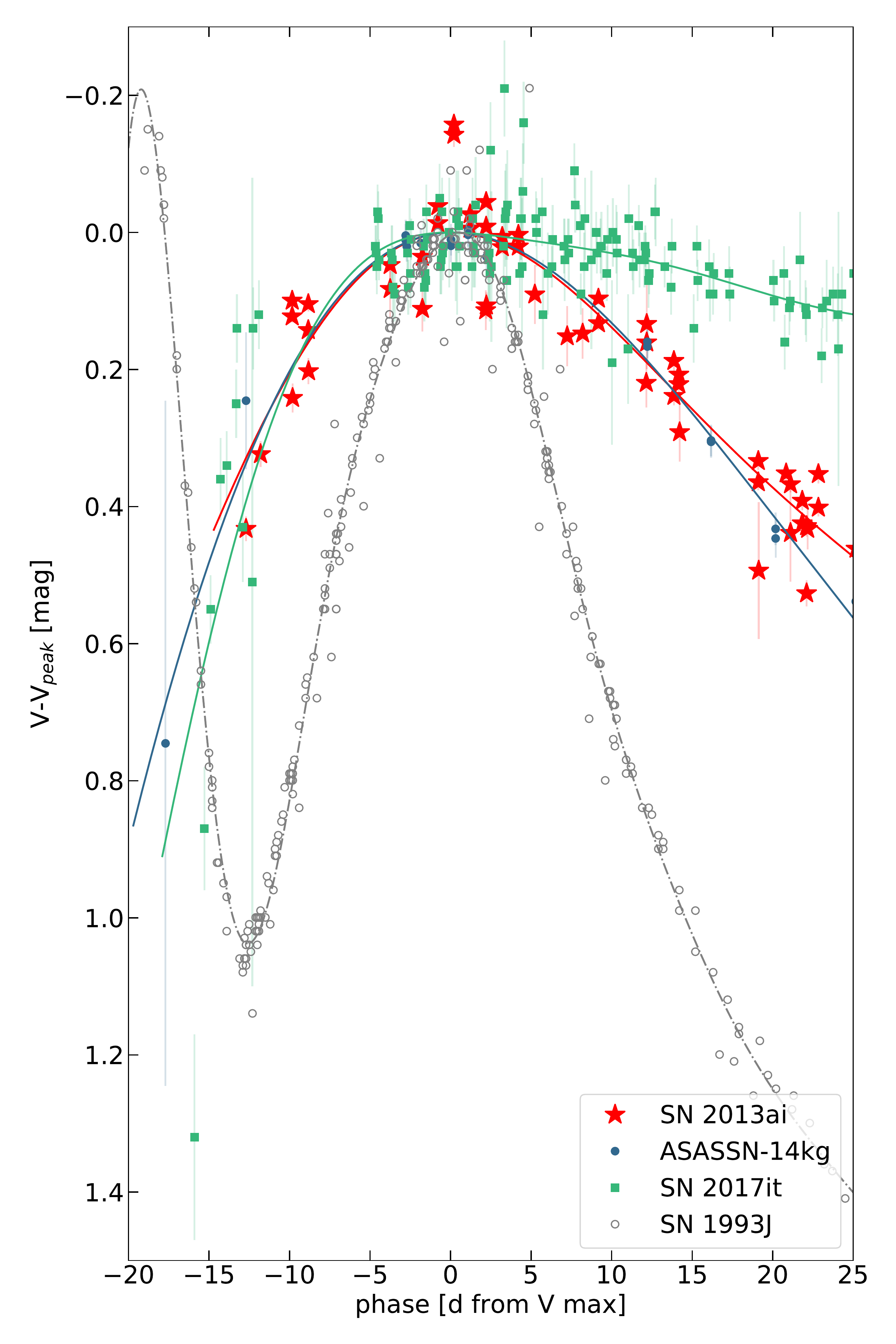}
    \caption{$V$-band light curves (dots) and Gaussian interpolations (lines) of SN~1993J, ASASSN-14kg, SN~2017it, and SN~2013ai normalized to peak magnitude. Light curves are without host subtractions. SN~1993J is included in order to compare with the longer rising and faster declining SN~IIb class.}
    \label{fig:13aiVS}
\end{figure}

SN~2013ai may be compared with two other SNe~II that were found in the literature: ASASSN-14kg \citep{Valenti2016} and SN~2017it \citep{Afsariardchi2019}. ASASSN-14kg was chosen as it was found to have the most similar light curve to SN~2013ai. SN~2017it was published as a high $^{56}$Ni mass SN~II with a long rise, so despite its slow decline post-maximum, bears comparison to SN~2013ai. Photometry from \citet{Valenti2016} shows a $V$-band rise time for ASASSN-14kg of $\sim$18 days while \citet{Afsariardchi2019} report a $V$-band rise time of $\sim$20 days for SN~2017it. In order to recalculate the rise times of these objects in a uniform fashion, the explosion epoch of the comparison SNe is taken as the mid point between the last non-detection and the first detection while considering the error to be half the difference between said epochs. The light curves were then interpolated via a Gaussian process method using the Python library {\sc GPy}\footnote{https://sheffieldml.github.io/GPy/}. Finally, the maximum date was obtained from the interpolated light curve and the rise times were calculated. Results are presented in Table \ref{tab:exp_epoch}.
It can be seen in Figure \ref{fig:13aiVS} that the behavior of SN~2013ai around peak is similar to ASASSN-14kg. SN~2017it, despite the long rise, exhibits a post-maximum decline similar to a normal SN~II. For further comparison to the SN~IIb subclass, SN~1993J, is added as an example of a typical SN~IIb light curve. The rise time of SN~2013ai is similar to a SN~IIb, however, the post-maximum decline is much slower.

To calculate the bolometric light curve, the Milky Way and NGC 2207 dust reddening corrected photometry was first converted into AB magnitudes using the spectral energy distribution of Vega. Monochromatic fluxes for each available date were numerically integrated in wavelength, obtaining the pseudo-bolometric flux (F$_{pbol}$). 
For the $UV$ flux (F$_{UV}$), a linear extrapolation was used and integrated from the observed flux at the $B$ band to zero flux at 2000~\AA\ \citep[see][for further details]{Bersten2009,Folatelli2014}. To estimate the IR flux (F$_{IR}$) the spectral energy distribution was fit by a grid of temperatures and angular sizes which define a black body for each point of the grid. Using the resultant $\chi^2$ grid from each black body, a probability distribution to determine the best parameters for each epoch is calculated. Afterwards, the flux was integrated from the effective wavelength of the $I$ filter to 100,000~\AA, where the flux is considered negligible.

\begin{figure}
    \includegraphics[width=\columnwidth]{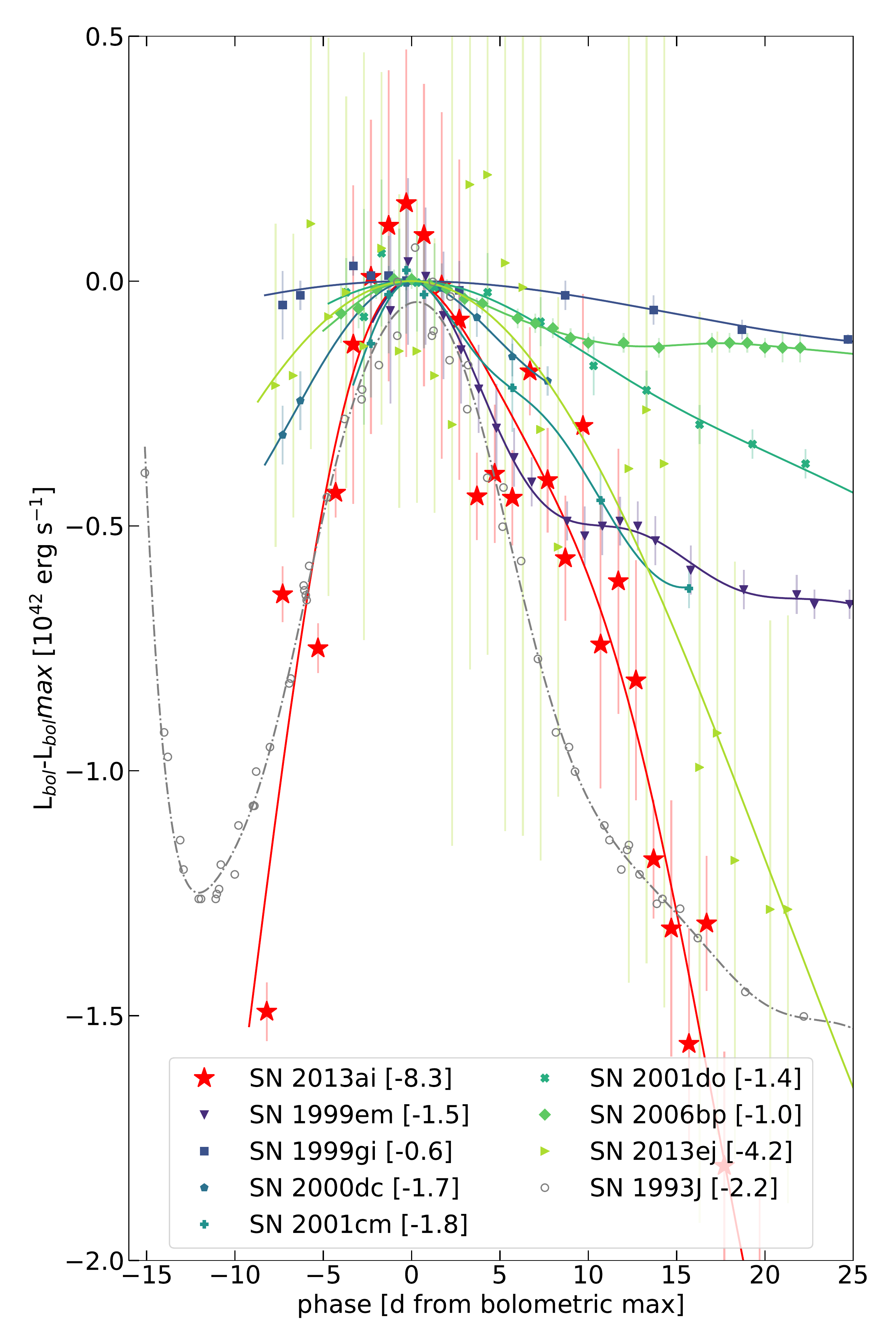}
    \caption{Bolometric light curve of SN~2013ai compared to a sub sample of \citet{Faran2018} objects, normalized by peak luminosity. The \citet{Faran2018} sample consists of normal SNe~II. The shift in luminosity, given in $10^{42}$~erg~s$^{-1}$, for each object is specified between brackets. The errorbars are taken from the bolometric corrections and are highly dependent on the amount of data taken for each epoch. SN~2013ai exhibits the steepest rise pre-maximum, among this slow rising sample, and quickest drop post-maximum. The bolometric light curve of SN~1993J is taken from \citet{Richmond1994} and is shown for comparison with the longer rising and faster declining SN~IIb class.}
    \label{fig:13aibol}
\end{figure}

The bolometric flux was calculated as $F_{bol}=F_{pbol}+F_{UV}+F_{IR}$. This bolometric light curve was compared to a subsample of the bolometric light curves presented by \citet{Faran2018}. Their sample includes only objects with high quality multiband data and rather good time coverage, which makes it useful for our goal of comparing SN~2013ai to a large sample of bolometric light curves computed in a uniform manner. 
Note that the bolometric light curves of \citet{Faran2018} were computed differently to the SN~2013ai bolometric light curves as they consider the effects in the bluer and redder wavelengths, e.g. line blanketing. However, given the uncertainties involved in calculating bolometric light curves, the comparisons presented would not significantly change if the bolometrics were recalculated. The subsample was selected such that only objects that show a clear maximum in the light curve, meaning there are data points before maximum, are considered. From the 29 objects presented in \citet{Faran2018}, only 8 meet this requirement, although SN~2003hf is much more luminous than the others so it was excluded, giving 7 comparison objects. The bolometric light curve of SN~1993J \citep{Richmond1994} is also presented for comparison. Figure \ref{fig:13aibol} shows that SN~2013ai is the fastest decliner, except for the SN~IIb SN~1993J. Using the explosion dates published by \citet{Faran2018} we obtain a mean bolometric rise time of the subsample of 8.6$\pm$2.0 days while the bolometric rise time of SN~2013ai is 14.3$\pm$4.1 days.

\begin{deluxetable*}{ccccccc}
\tablenum{4}
\tablecaption{Explosion epoch for comparison objects.}
\centering
\label{tab:exp_epoch}
\tablecolumns{7}
\tablehead{
\colhead{SN}           &
\colhead{Last Non-detection} &
\colhead{First Detection}    &
\colhead{Explosion Epoch}  &
\colhead{Reference}         &
\colhead{$V$-band Max Epoch} &
\colhead{$V$-band Rise Time}
}
\startdata
ASASSN-14kg & 56972.4 & 56973.5  & 56972.9$\pm$0.5  & \citet{atel6714} & 56991.1$\pm$2.8  & 18.1$\pm$2.9    \\
SN~2013ai   & 56344.3 & 56349.7  & 56347.0$\pm$2.7 & \citet{atel4849} & 56365.9$\pm$2.6  & 18.9$\pm$3.7    \\
SN~2017it   & 57745.5 & 57746.9	 & 57746.2$\pm$0.7  & \citet{Afsariardchi2019} & 57762.8$\pm$4.2  & 16.6$\pm$4.3
\enddata
\tablecomments{All dates are given in MJD.}
\end{deluxetable*}

\section{Spectral Properties}
\label{spec}

The optical and NIR spectra of SN~2013ai were analyzed quantitatively by measuring expansion velocities and pEWs for every feature present. These techniques have been implemented succesfully before to study large samples of SNe II and their diversity \citep[e.g.][]{Gutierrez2017spec,Davis2019,deJaeger2019}.

\subsection{Time Evolution}

Figure \ref{fig:optSpec} shows the optical spectra of SN~2013ai. In the optical, at all epochs, we see narrow optical emission in H$\alpha$, H$\beta$, [O~{\sc ii}] $\lambda$3727, [O~{\sc iii}] $\lambda$5007 and [S~{\sc ii}] $\lambda\lambda$6717, 6731 at the redshift of NGC 2207. The earliest spectra, taken $11$ days before $V$ maximum, show only a broad H$\alpha$ P Cygni profile with narrow Na~I~D $\lambda\lambda$5890 and 5896 likely from the interstellar medium (ISM). H$\beta$ is not present until later times, first seen around $3.2$ days before $V$ maximum. A montage of the optical spectra is presented in Figure \ref{fig:optSpec}.

Starting with the spectrum taken $3$ days after $V$ maximum, the H$\alpha$ absorption becomes split, likely due to Si~{\sc ii} $\lambda$6355 at these early times \citep{Gutierrez2017spec}. Fe-group lines are also present during this epoch, primarily Fe~{\sc ii} $\lambda\lambda$5169, 5267, and 5363. We also see the emergence of the Ca~{\sc ii} NIR triplet as a broad and blended P Cygni line around 8500~\AA. 

From $11-20$ days after maximum, SN~2013ai became even redder, as more flux is lost to absorption by Cr, Sc, Ba and Fe in the blue. The usual Fe-group lines seen in SNe~II are present between 3500-5500~\AA; however, these lines are weaker than what are seen in a normal SN~II (see \citealt{Gutierrez2017spec,deJaeger2019}). H$\alpha$ has a strong absorption component in its P Cygni profile, with a minimum at a velocity of 11,000 \kms, while the second absorption, which was tentatively associated with Si~{\sc ii}, is still present as a ``notch" in the feature. There is a lack of absorption from Sc~{\sc ii} and Ba~{\sc ii} on the blue side of the H$\alpha$ absorption line.

By $30$ days after maximum, the notch in H$\alpha$ is no longer present in the spectrum of SN~2013ai, suggesting that the notch was not due to high velocity (HV) H~{\sc i} (see \citealt{Gutierrez2017spec} for more information on this feature). The spectrum shows relatively little evolution over the following period, from $30$ to $73$ days post explosion, the continuum continues to become redder, while the most notable change in the lines is an increase in the strength of the Na~I~D absorption.

\begin{figure*}[t]
    \includegraphics[width=\textwidth]{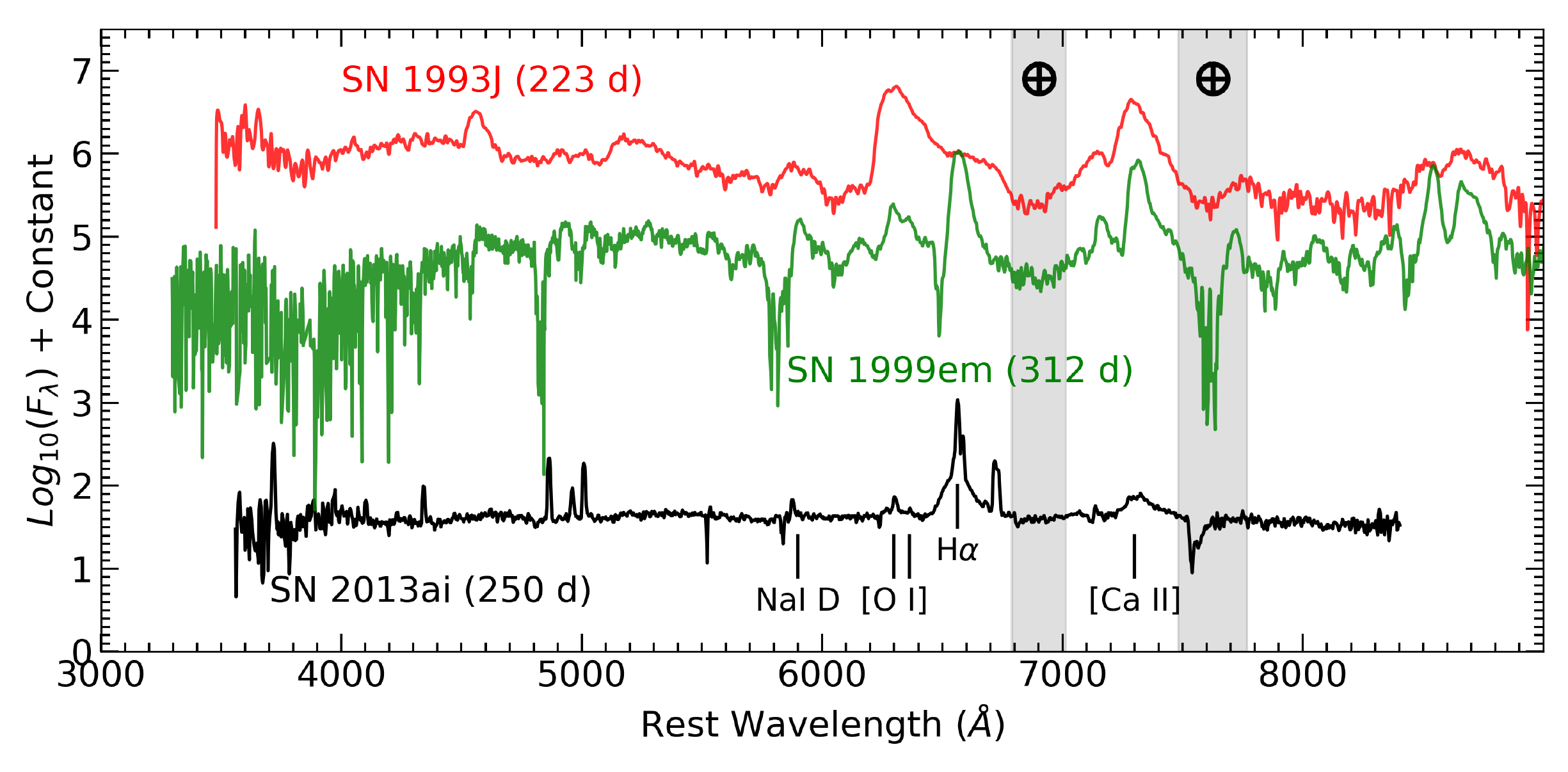}
    \caption{The late time nebular spectrum of SN~2013ai taken with GTC+OSIRIS on 2013 October 30 (black), compared to spectra of the Type IIP SN~1999em and the Type IIb SN~1993J at a comparable epoch (from \citealt{Elmhamdi2003} and \citealt{Matheson2000} respectively). The spectrum has contamination from the host seen primarily around 5000 \AA\ in emission. The features present in the spectrum of SN~2013ai are identified, while the regions of strong telluric absorption are indicated with a $\oplus$ symbol. The H$\alpha$ profile is two-tiered and has no absorption at this phase, suggesting interaction with CSM.}
    \label{fig:late_spec}
\end{figure*}

The late time optical spectrum obtained $230$ days after $V$ maximum with GTC+OSIRIS is shown in Figure \ref{fig:late_spec}. This spectrum was taken 250 days from the estimated explosion epoch and is shown in comparison to the SN~IIP SNe~1999em and the SN~IIb SN~1993J at similar times. 
The spectrum of SN~2013ai contains very little flux from the SN. Nonetheless, several lines characteristic of core-collapse SNe in their nebular phase are detected, namely broad H$\alpha$ emission, [Ca~{\sc ii}] $\lambda$7291, and possibly [O~{\sc i}] $\lambda$6300. 
The reddening corrected [Ca~{\sc ii}] to [O~{\sc i}] line ratio appears approximately similar to that in SN~2004et \citep[see][]{Jerkstrand2012}, suggesting that the core mass of the star which exploded was probably quite similar and not exceptionally massive. However, the weakness of these lines makes it difficult to accurately measure a flux ratio and suggest that the SN is not sufficiently nebular for an accurate progenitor to be assumed using the method of \citet{Jerkstrand2012}.
The H$\alpha$ profile at this epoch is dominated by emission and shows a profile with a wide base and narrow top with no absorption. Together with the early X-ray observations, as noted in Section \ref{obs}, this suggests interaction with circumstellar material (CSM). 

\begin{figure}
    \includegraphics[width=\columnwidth]{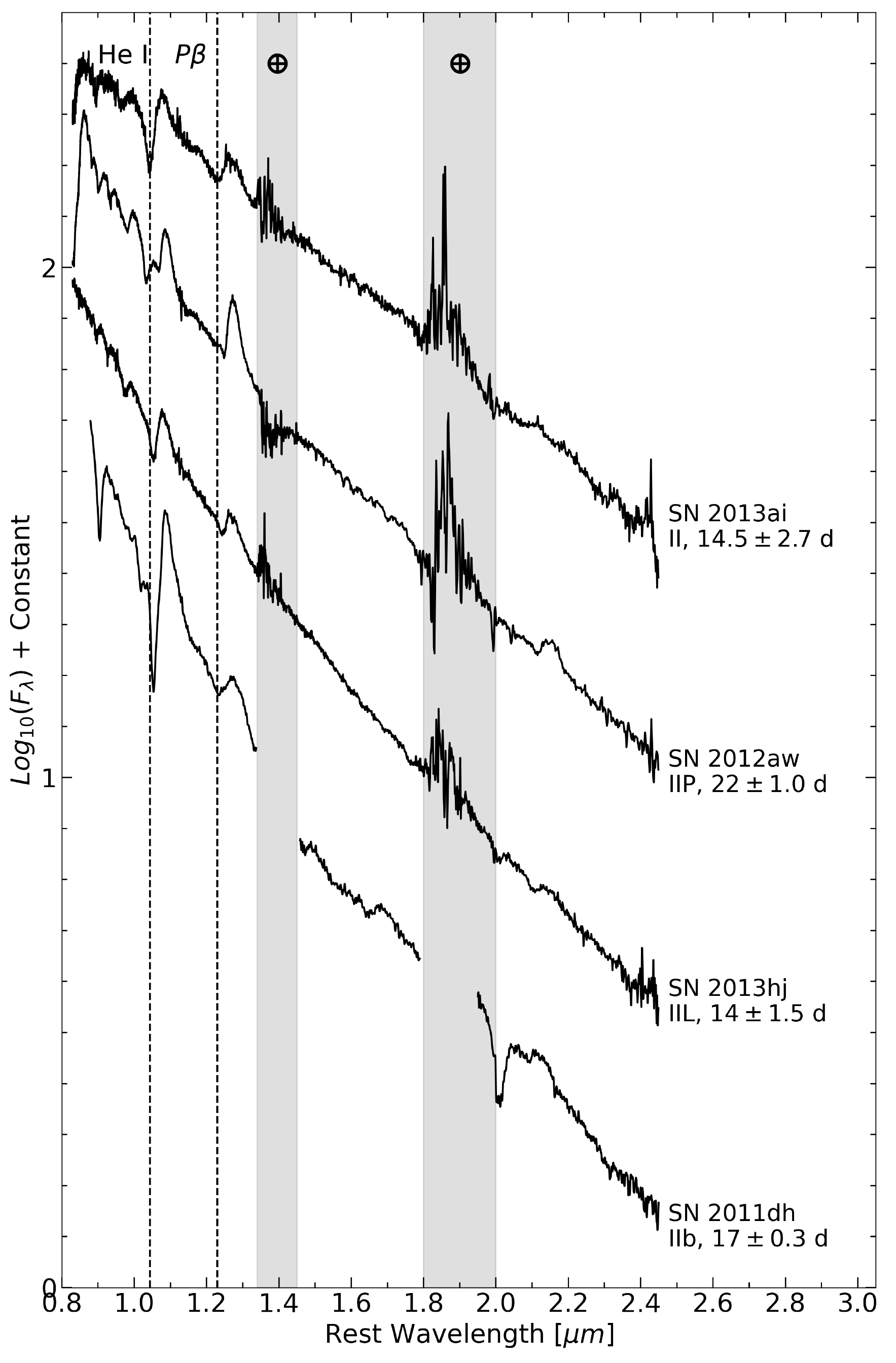}
    \caption{NIR spectra comparison of SN~2013ai with other SNe II around the same epoch. Prominent H and He line absorptions are marked with dashed vertical lines to emphasis the fast velocities of SN~2013ai. The phase from explosion is listed for each spectrum along with the SN subtype. The gray vertical bands mark the regions of strongest telluric absorption. SN~2011dh has little data taken within the telluric bands and it is not plotted in these regions. SN~2013ai shows no sign of being a weak SN~II \citep{Davis2019}. The spectra of SNe~2012aw and 2013hj are from \citet{Davis2019}. The spectrum of SN~2011dh is from \citet{Ergon2014}.}
    \label{fig:NIRspec}
\end{figure}

The NIR spectrum of SN~2013ai is shown in Figure \ref{fig:NIRspec} and exhibits mostly hydrogen and helium features, such as $P_{\delta}$ and $B_{\gamma}$ and the He~{\sc i}/$P_{\gamma}$ blend. In the NIR, SN~2013ai does not exhibit any lack of features like the absence Sc~{\sc ii} and Ba~{\sc ii} in the optical.

\subsection{Comparison to Other Core-Collapse SNe}

\begin{figure*}
    \includegraphics[width=\textwidth]{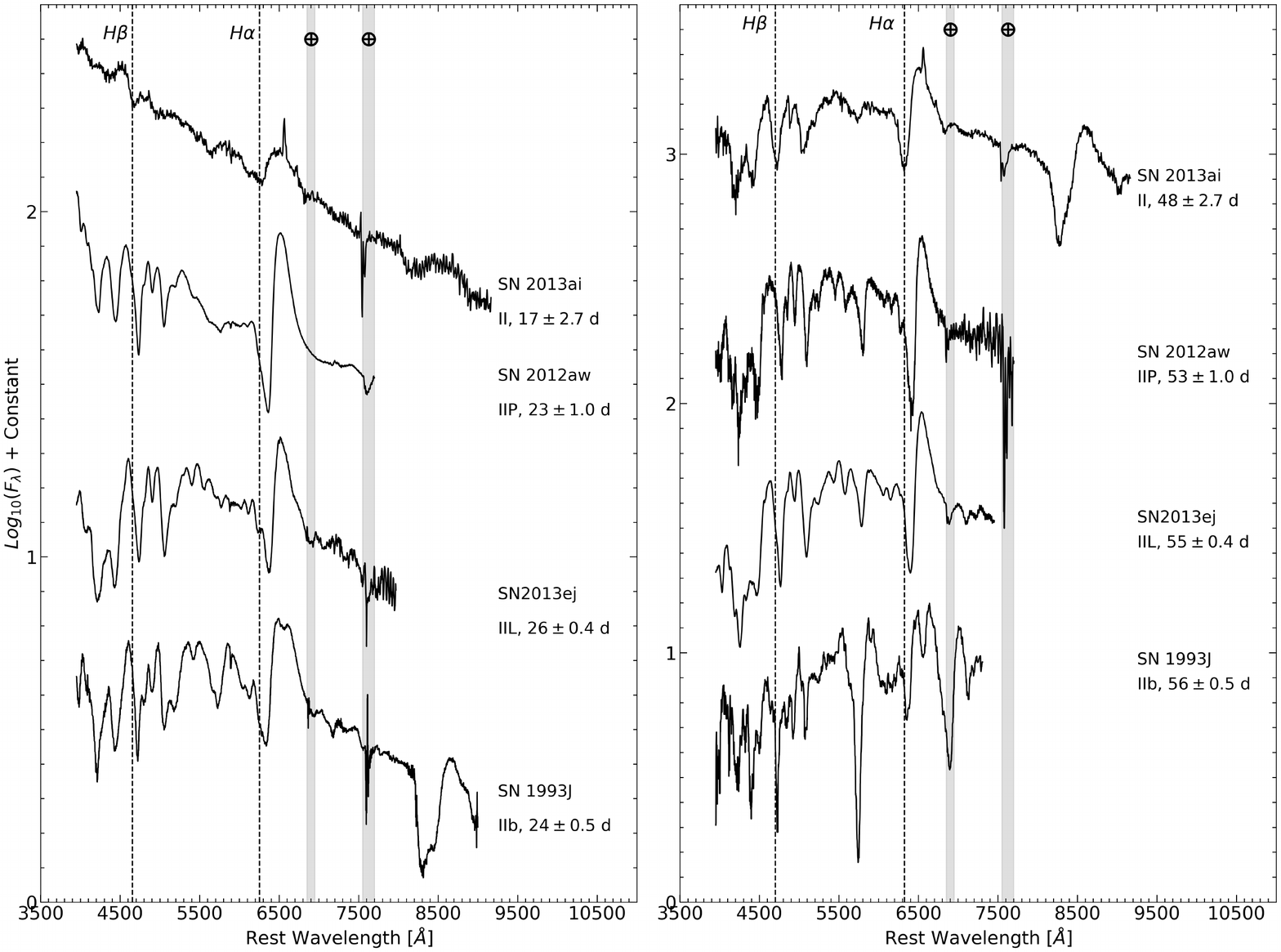}
    \caption{Optical spectra comparison of SN~2013ai with other SNe II during the photospheric phase. Prominent H line absorptions are marked with dashed vertical lines to emphasise the fast velocities of SN~2013ai. The phase from explosion is listed for each spectrum along with the SN subtype. The gray vertical bands mark the regions of strongest telluric absorption. Data from SNe~2012aw, 2013ej, and 1993J are from \citet{DallOra2014}, \cite{Childress2016}, and \citet{Barbon1995}, respectively.}
    \label{fig:optCompareComb}
\end{figure*}

Figure \ref{fig:optCompareComb} shows optical spectra of SN~2013ai compared to other SNe at similar phases. Included are examples of well-studied SNe~IIP, IIL, and IIb. This comparison highlights the uniqueness of SN~2013ai in the features seen, their strengths, and expansion velocities. As previously noted, there is a lack of features on the blue side of H$\alpha$ and the Na/He absorption around 5990~\AA\ is significantly shallower than that seen in a normal SN~II/IIb.

Figure \ref{fig:NIRspec} shows the 14.5 day NIR spectrum of SN~2013ai compared to other SNe II at similar epochs. The SN~does not show any signs of being a NIR \emph{weak} SN~II \citep{Davis2019}; however, at such early times it is not always possible to determine a NIR spectroscopic subclass. The He~{\sc i} $1.083\,\mu m$ line is stronger than that of SN~2013hj at a similar phase. Both SNe show a boxy $P_{\beta}$ emission profile. Overall, the NIR features of SN~2013ai are much more normal than in the optical. However, similar to the optical, the features in the NIR have much higher velocities than other SNe IIP/L at similar times.



\begin{figure*}[t]
    \includegraphics[width=\textwidth]{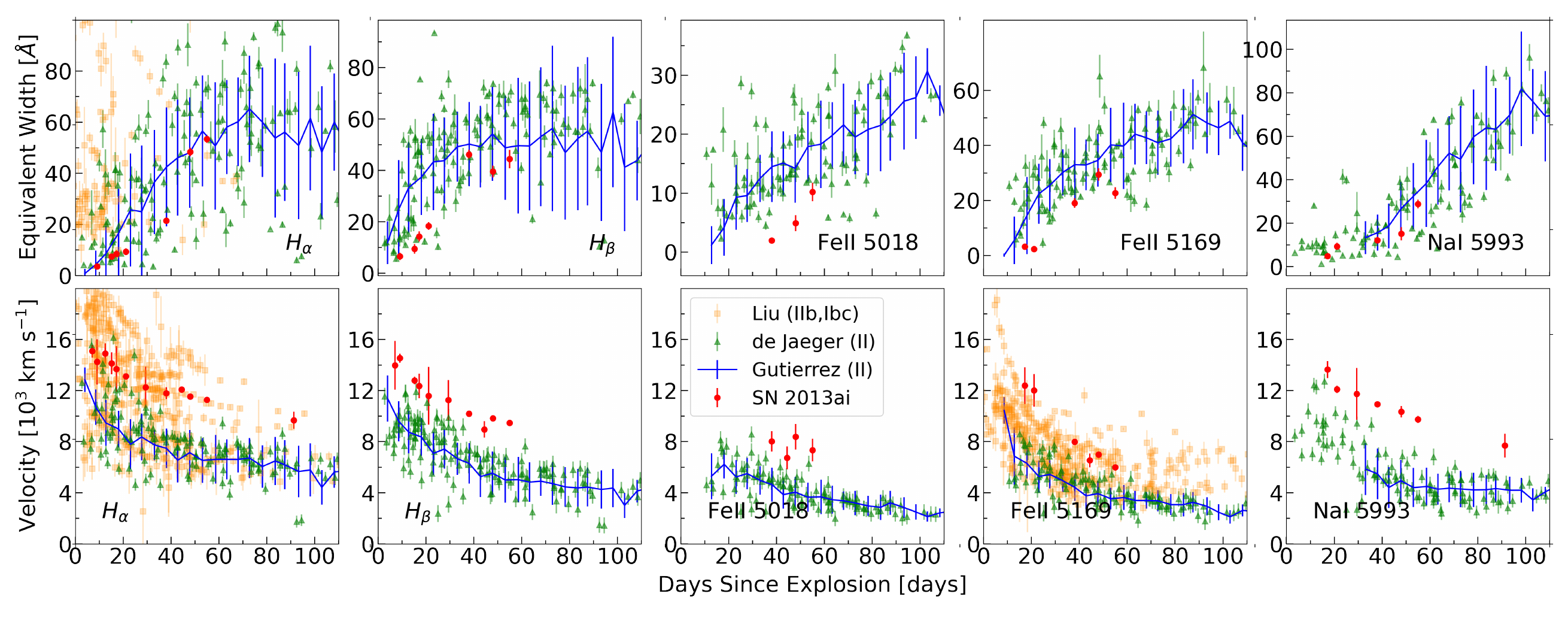}
    \caption{Comparison of SN~2013ai optical absorption pEWs and velocities over time to the mean values from \citet{Gutierrez2017spec} and all data from \citet{Liu2016} and \citet{deJaeger2019}. 
    The samples of \citet{Gutierrez2017spec} and \citet{deJaeger2019} are made up of SNe~II. Included is data from \citet{Liu2016} in order to compare to SESNe. Data from \citet{Liu2016} was published relative to maximum, thus for this comparison we determine the explosion date for each SN in the \citet{Liu2016} sample from the last non-detection method described in Section \ref{obs}. SN~2013ai matches more closely to the velocities of a SESN than a SN~II.
    The velocities of all features of SN~2013ai start much higher ($\sim2 \sigma$) and stay high while decreasing at a similar rate to normal SNe~II. The pEWs of SN~2013ai match within 1 sigma of the mean. No SN in either SN~II sample sustains velocities as high as SN~2013ai.}
    \label{fig:Gcompare}
\end{figure*}

\begin{figure*}[t]
    \includegraphics[width=\textwidth]{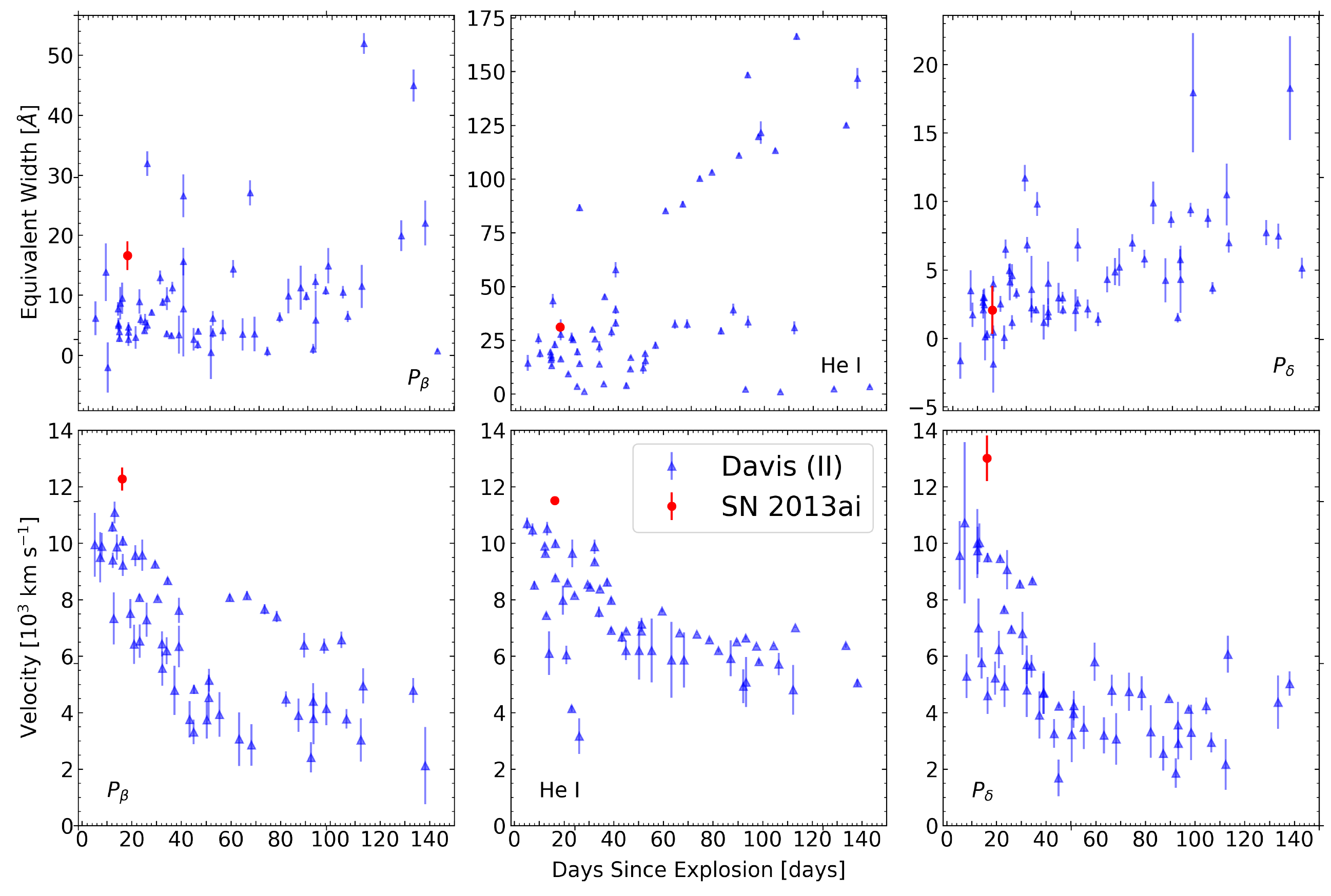}
    \caption{SN~2013ai (large red dots) compared to the sample of normal SNe~II from \citet{Davis2019} (blue triangles) in NIR pEWs and velocities. SN~2013ai is at significantly higher velocities than any SN in the \citet{Davis2019} sample. However, much like the optical, SN~2013ai has normal pEWs when compared with SNe~II.}
    \label{fig:NIRcompare}
\end{figure*}

The top row of Figure \ref{fig:Gcompare} compares the pEW of all optical features present in the spectra of SN~2013ai with the mean from \citet{Gutierrez2017spec} and all data from the \citet{deJaeger2019} and \citet{Liu2016} samples. 
The metal lines, from 4500-5500~\AA, of SN~2013ai evolve like those of a typical SN~II. 
The 5990~\AA\ He/Na blend is particularly interesting due to its weakness at later times. The pEWs of all features seen in the optical lie amongst normal SNe~II values. Similarly, the top row of Figure \ref{fig:NIRcompare} compares the pEW of all NIR spectral features of SN~2013ai with the sample from \citet{Davis2019}. The NIR pEWs, like the optical, lie within the large comparative samples.

The bottom row of Figure \ref{fig:Gcompare} shows the absorption velocities of SN~2013ai compared to the samples of \citet{Liu2016}, \citet{Gutierrez2017spec}, and \citet{deJaeger2019}. SN~2013ai exhibits high velocities in all features observed, for a SN~II.
The 5990~\AA\ He/Na blend is particularly interesting due to its high velocities if assumed to be Na~{\sc i} $\lambda5993$, giving velocities over 13,000 \kms\ at early  times. While the Fe~{\sc ii} velocities are closer to normal SN~II values, they are still noticeably higher than the comparative SN~II samples, with the velocities more similar to SESNe. Despite the high velocities, the evolution is decreasing with time, as that of a normal SN~II. Figure \ref{fig:NIRcompare} shows the NIR absorption velocities of SN~2013ai compared to that of the data from \citet{Davis2019}. Much like the optical, the velocities are significantly higher than the sample, $\sim2\sigma$, with the P$\beta$ and P$\delta$ features around 12,000-13,000 \kms\ and He~{\sc i} $\lambda1.083\,\mu m$ around 11,500 \kms at 22 days past maximum. The offset of velocities from typical SN~II values are similar in the optical and NIR. The normal pEWs and high velocities of SN~2013ai suggest that the features are not broad, but are fast, like those of a SN~IIb. However, the spectral features present are typical of a SN~II.

The optical absorption velocities at 50 days past explosion, the middle of the plateau for a normal SN~II, were also compared to the \citet{Gutierrez2017spec} and \citet{deJaeger2019} samples in order to see if any other SNe II have been found with velocities as high as SN~2013ai.
No SNe with velocities as high as SN~2013ai were found, except for SN~2007ld from the \citet{Gutierrez2017spec} sample, which has a 50 day H$\alpha$ velocity over 11,000 \kms.
SN~2007ld has a normal photometric and spectroscopic evolution; however, it shows some contamination in the H$\alpha$ emission profile which could be due to He~{\sc i}, a common sign of a SN~IIb. Unfortunately, SN~2007ld has no spectra towards the end of its plateau, when a SN~IIb can be easily identified. SN~2007ld does not exhibit a long rising light curve or a quick decline after maximum. 
Super-luminous SNe \citep[e.g. SN~2013fc, SN~2016gsd][]{Kangas2016,Reynolds2020} are not included for comparison as SN~2013ai is not super-luminous. However, see SN~2016gsd \citep{Reynolds2020} for a super-luminous SNe~II that exhibits high velocities similar to SN~2013ai.

\begin{figure*}[t]
    \includegraphics[width=\textwidth]{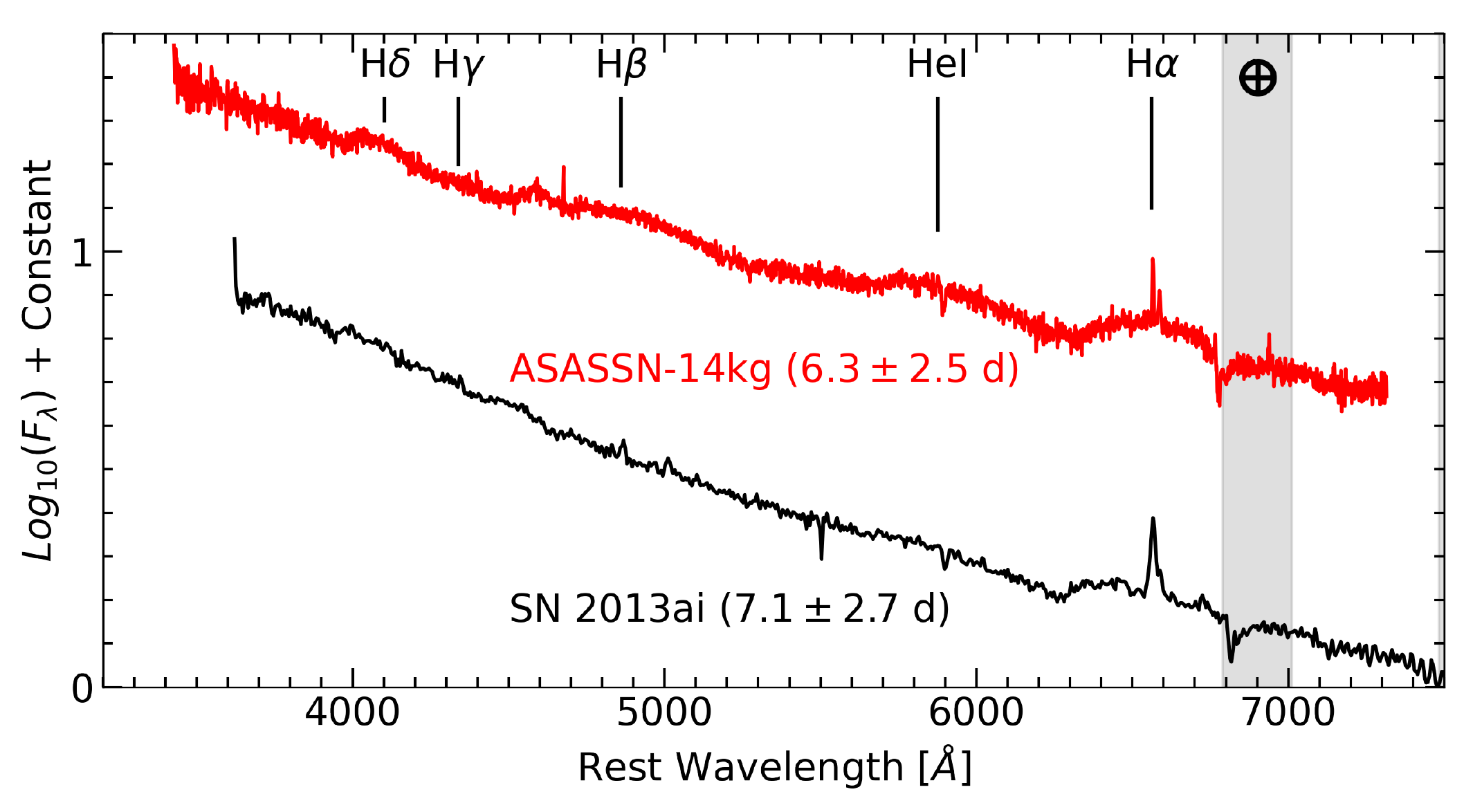}
    \caption{Comparison of the early spectrum SN~2013ai to ASASSN-14kg. The spectra are dereddened. Both SNe show high velocities and have similar spectra at these early phases. The rest wavelength of each hydrogen and helium line possibly present are labeled. Regions of strong telluric absorption are outlined in gray. ASASSN14-kg is the only SN found to be similar to SN~2013ai in both spectral features and velocities. }
    \label{fig:14kg}
\end{figure*}

The spectra of SNe with similar rise times to SN~2013ai (see Section \ref{sec:phot}) were also examined. 
The classification spectrum of ASASSN-14kg is plotted with the earliest spectrum of SN~2013ai in Figure \ref{fig:14kg}. The spectra look similar at these early times. There are no later spectra taken of ASASSN-14kg. 
ASASSN-14kg has a well defined explosion date, $59672.9 \pm 0.5$ MJD \citep{atel6714}. Given the similarity of the light curves and early spectra of ASASSN-14g and SN~2013ai, it is likely that SN~2013ai is around the same phase from explosion in the spectra seen in Figure \ref{fig:14kg}.
\citet{Afsariardchi2019} presented the H$\alpha$ velocity of SN~2017it at 93 days past explosion, which is higher than that of a normal SNe~II. However, it is still $\sim$1000 \kms\ slower than SN~2013ai. We were unable to find any SN~II with expansion velocities comparable to SN~2013ai during the photospheric phase. However, the similarity of the early-time spectra and light curves of ASASSN-14kg and SN~2013ai suggests that they could be alike. 

\section{Progenitor}
\label{sect:progenitor}

\subsection{Pre-explosion Observations}
\begin{figure*}
\gridline{\fig{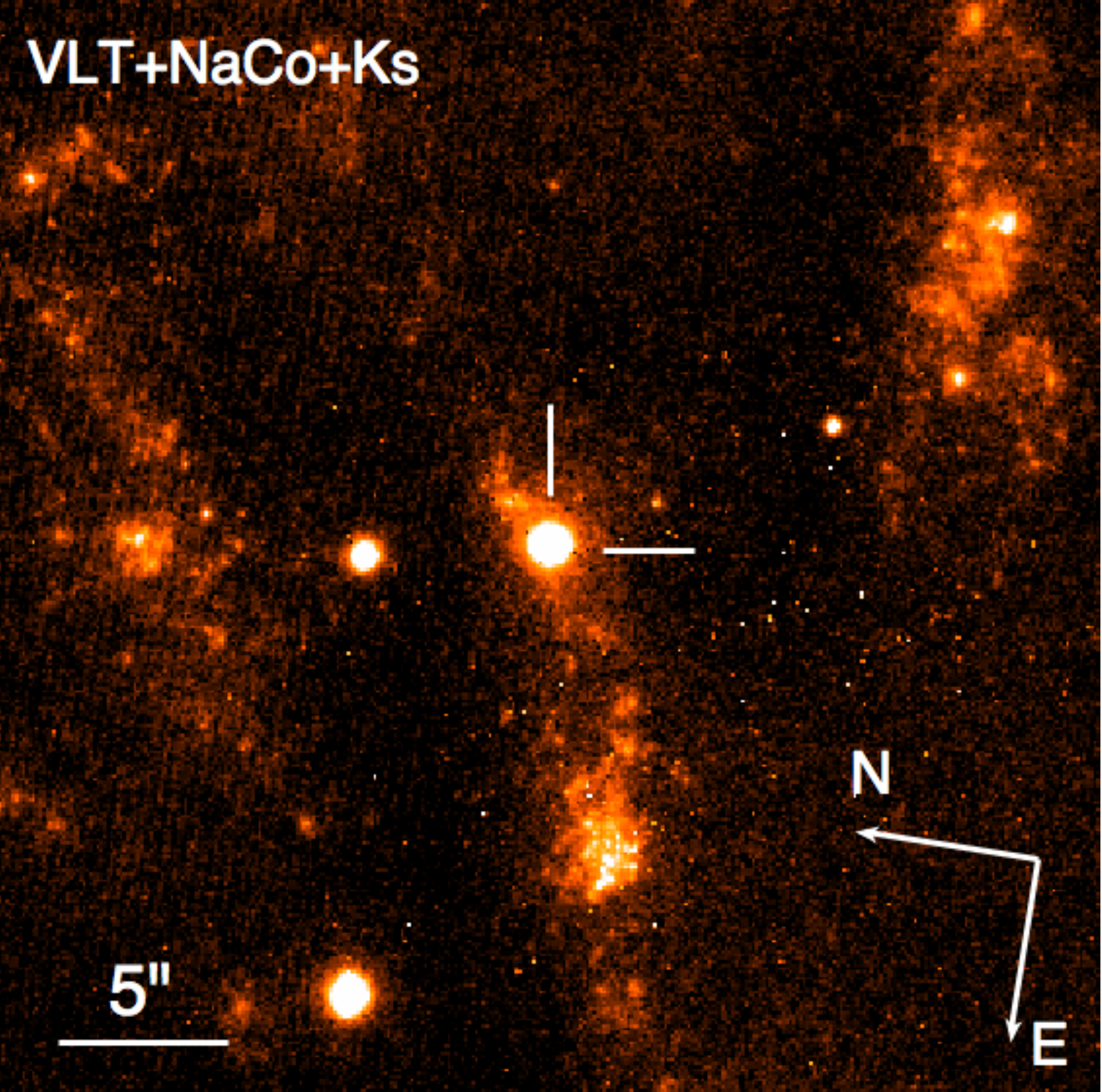}{0.3\textwidth}{(a)}
          \fig{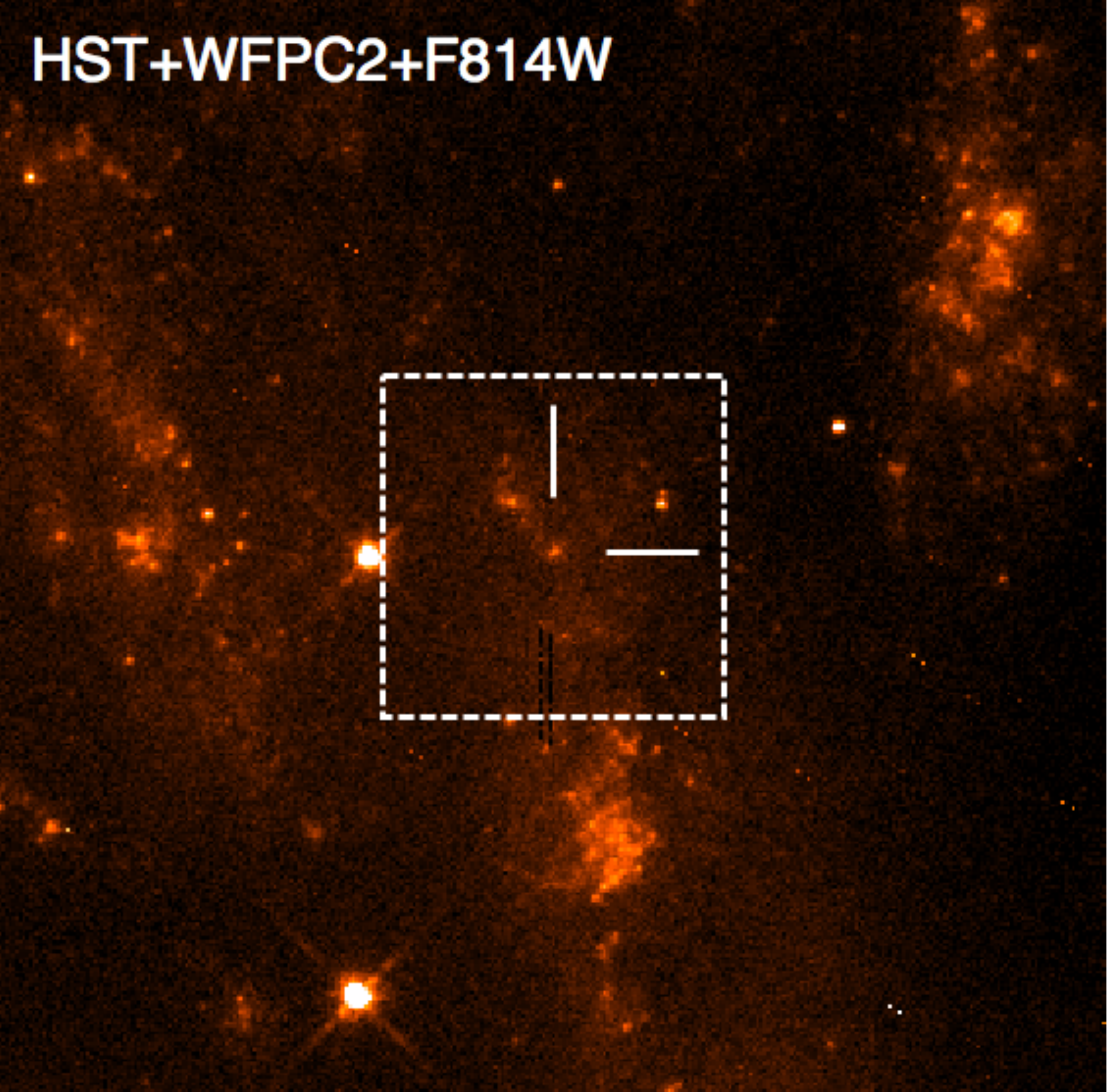}{0.3\textwidth}{(b)}
          \fig{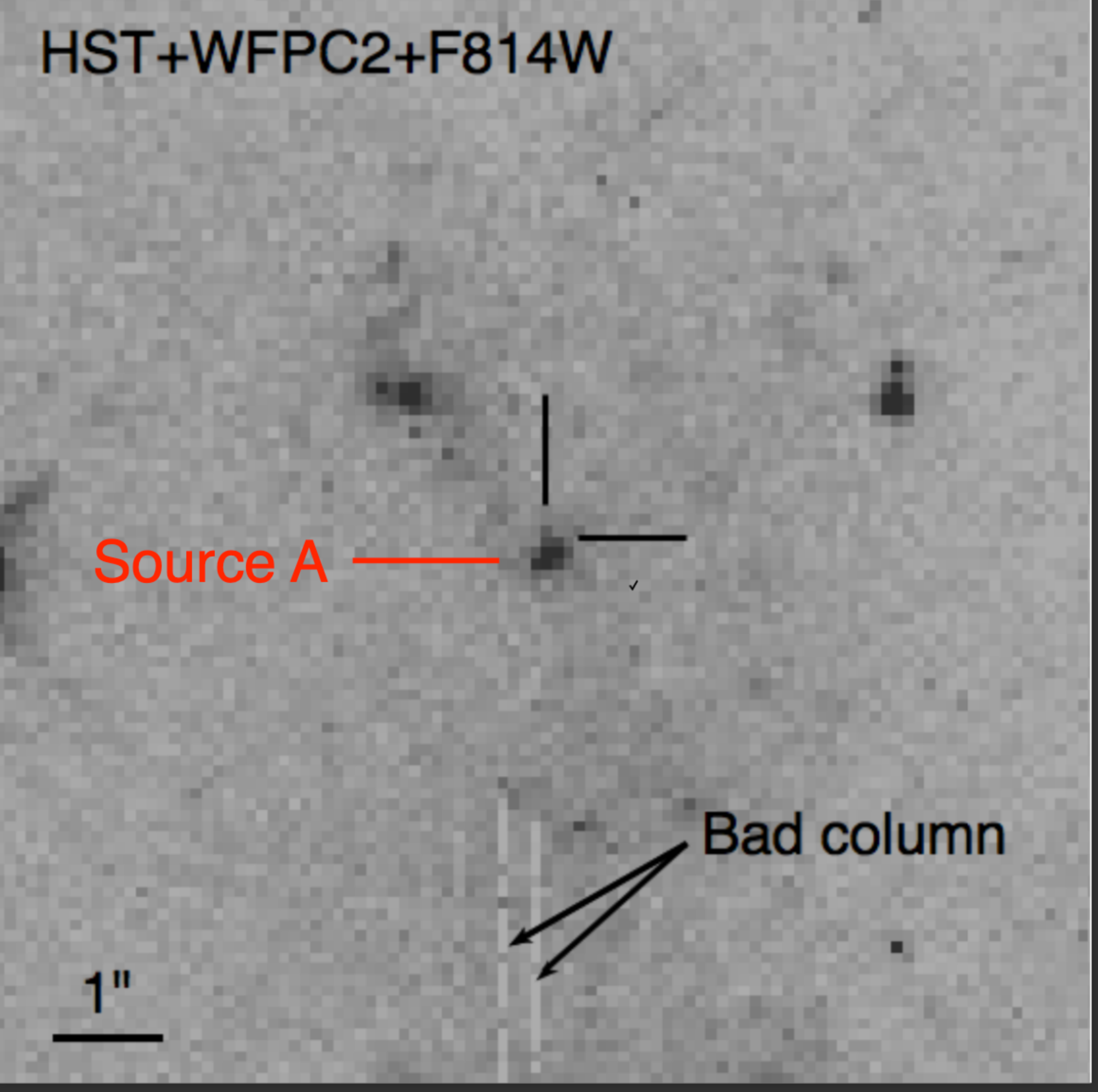}{0.3\textwidth}{(c)}}
\caption{Pre- and post-explosion images of SN~2013ai from HST+WFPC2 and VLT+NaCo. (a) VLT {\it Ks} image of SN~2013ai. (b) Section of HST WFPC2 F814W pre-explosion image, with the same orientation and scale as in panel (a). The dashed box is shown in more detail in panel (c). (c) Region indicated with the dashed box in panel (b). The transformed SN~pixel coordinates are indicated with cross hairs, while the bad column discussed in the text is also indicated.}
\label{fig:progenitor}
\end{figure*}

Pre-explosion HST observations were available for SN~2013ai, consisting of HST+Wide-Field and Planetary Camera 2 (WFPC2) images taken on 1996 May 25, see Figure \ref{fig:progenitor}. The SN location lies on the WF2 chip, which has a pixel scale of 0.1\arcsec/pixel. Four exposures, consisting of two cr-split pairs, were taken in each of the F336W, F439W, F555W, and F814W filters, with total exposure times of 2000s, 2000s, 660s, and 720s, respectively. The pipeline reduced {\tt \_c0f} files were downloaded from the Mikulski Archive for Space Telescopes (MAST)\footnote{http://archive.stsci.edu/}. The images were first masked with their associated {\tt \_c1f} files, and the cr-split pairs were combined to reject cosmic rays using the {\sc crmask} task within {\sc IRAF}. Finally, the dithered images in each filter were aligned and coadded.

Unfortunately, the SN position lies very close to the charge transfer trap 2-337 on the WF2 chip in all images \citep{Whitmore1995}, which gives rise to a bad column artefact. While it is possible to attempt to reconstruct the flux lost due to the trap, as dithered exposures were available, we instead ensured that the bad column was masked in both cr-split pairs for each filter before coadding. As the cr-split pairs in each filter were offset by $\sim$3 pixels in the x direction, the defect (when masked) does not contribute to the final image when they are shifted and combined.

We also attempted to use the {\sc drizzle} algorithm \citep{Fruchter1997} as implemented in {\sc astrodrizzle} within the {\sc drizzlepac} package to improve the spatial resolution of the F814W image. Before drizzling, the bad column close to the position of SN~2013ai was flagged in the data quality image. A series of drizzled images were produced, to test the effect of varying the output pixel scale and the size of the drop. However, in all cases a faint artefact was still visible at the position of SN~2013ai in the final image. It is likely that the artefact in the output image results from the interpolation over the masked bad column. Thus the drizzled images were not considered any further.

To identify the position of SN~2013ai in our pre-explosion images, we obtained {\it Ks} filter imaging of SN~2013ai with the Nasmyth Adaptive Optics System and Near-Infrared Imager and Spectrograph (NAOS-CONICA; NaCo) on the Very Large Telescope UT4 over 2013 March 3-4. SN~2013ai itself was used as a guide star for NAOS, while the S54 camera was employed on CONICA, yielding a pixel scale of 0.0543\arcsec / pixel, over a field of view of 56\arcsec$\times$56\arcsec. Daytime calibration data consisting of flat fields and long and short exposure darks were reduced with the NaCo pipeline (version 4.3.1), to give a master flat field and a map of aberrant pixels on the detector. Using these, the individual science frames were masked and flat-fielded. All images of SN~2013ai were dithered on-source, facilitating the subtraction of the sky background using the {\sc IRAF xdimsum} package. After sky-subtraction, the individual frames were aligned and combined to yield a single, deep image with a total exposure time of 5040s.

The NaCo post-explosion image was aligned to the pre-explosion WFPC2 F814W image using a geometric transformation derived from the pixel coordinates of sources common to both images. Two separate sets of sources were used for the alignment: a ``good'' set of 21 sources which were clearly point-like, detected with good S/N, and which appear to lie within NGC 2207, and a larger superset of 31 sources which we term the ``complete'' set, where the signal to noise (S/N) of sources were lower, and which also included foreground and slightly extended objects. For both the ``good'' and ``complete'' sets, the pixel coordinates of each source were measured in the WFPC2 and NaCo frames, and three separate geometric transformations were then derived using the lists of matched coordinates, one allowing for translation, rotation, and scaling in x and y, and two which also included second and third order polynomial terms, respectively.

The pixel coordinates of SN~2013ai were measured in the NaCo image using three different algorithms (centroid, gauss and ofilter within {\sc IRAF}), which had a standard deviation of only 0.05 pixels, or 3 mas. The average of the three algorithms was taken as the position of SN~2013ai in the NaCo image. This position was then transformed to the pixel coordinates of the pre-explosion image using each of the transformations derived using both the ``good'' and ``complete'' sources. The standard deviation of the transformed positions was 16 mas, and we take this to be indicative of the uncertainty in SN position depending on the geometry or sources used for the alignment. We add this in quadrature with the uncertainty in the SN~position in the NaCo observation (3 mas), and the average of the RMS error found when fitting each transformation (58 mas), to obtain the total uncertainty of 60 mas on the location of SN~2013ai in the WFPC2 image.

\subsection{Progenitor Age Estimate} 

The SN~falls close to, but is not coincident, with a source in the WFPC2 image, which we designate ``Source A''. The centre of Source A (see Fig. \ref{fig:progenitor}), as measured using three centering algorithms within {\sc IRAF phot}, lies 193 mas (1.93 pixels) from the transformed SN~position. This offset is a factor $\sim$3 greater than the uncertainty in the transformed position of SN~2013ai (60 mas) and so we discount the possibility that this is a single stellar progenitor of SN~2013ai.

By eye, Source A appears to be somewhat extended. To test this, 14 isolated, point-like sources were fit with Moffat profiles to determine their full-width at half maximum (FWHM). The average FWHM of these sources was 1.6$\pm$0.2 pixels. Source A is significantly broader, with a FWHM of 2.5 pixels. On these grounds, it appears that the source may be an unresolved cluster or complex (2.5 WF pixels corresponds to a physical scale of 46 pc at the distance of NGC 2207).

While the position of SN~2013ai lies outside the FWHM of Source A, it still appears to fall within the wings of the PSF. To quantify the degree to which SN~2013ai lies within the position of Source A, the latter was fit with a Moffat profile. Then, we integrated under this profile to determine the flux within an aperture with a radius corresponding to the distance from Source A to SN~2013ai. Assuming that SN~2013ai is associated with a cluster, it was found that 97$^{+2}_{-6}$ percent of the flux of Source A is in the position of SN~2013ai. 
In the remainder of this section, we consider the implications for the progenitor of SN~2013ai both if it was associated with Source A, and if it was unrelated. 

Photometry was performed on Source A using {\sc hstphot}, \citep{Dolphin2000a} a stand-alone photometry package for use with WFPC2 data. {\sc hstphot} includes corrections for charge transfer efficiency (CTE) losses, aperture corrections, and zeropoints appropriate to WFPC2 \citep{Dolphin2000b}. While Source A is somewhat broader than most point sources in the field, it is still relatively well fit by a PSF. We measure PSF-fitted magnitudes for Source A in the HST flight system \citep{Holtzman1995}, with updated zeropoints as per Andrew Dolphin's webpages\footnote{http://americano.dolphinsim.com/wfpc2\_calib/} of 
F336W=22.68$\pm$0.12~mag,
F439W=23.90$\pm$0.15~mag,
F555W=23.40$\pm$0.12~mag, 
F814W=22.60$\pm$0.13~mag.
Transforming to the standard {\it UBVRI} system, these correspond to 
{\it U}=22.68$\pm$0.12~mag,
{\it B}=23.89$\pm$0.15~mag,  
{\it V}=23.38$\pm$0.12~mag,
{\it I}=22.56$\pm$0.13~mag. 
As a test of the PSF-fitting results, aperture photometry was also performed using a small (2 WF pixel) aperture. For all four filters, the difference between PSF-fitting and aperture photometry was $<$0.1 mag, i.e. less than the photometric error.

\begin{figure*}
\gridline{\fig{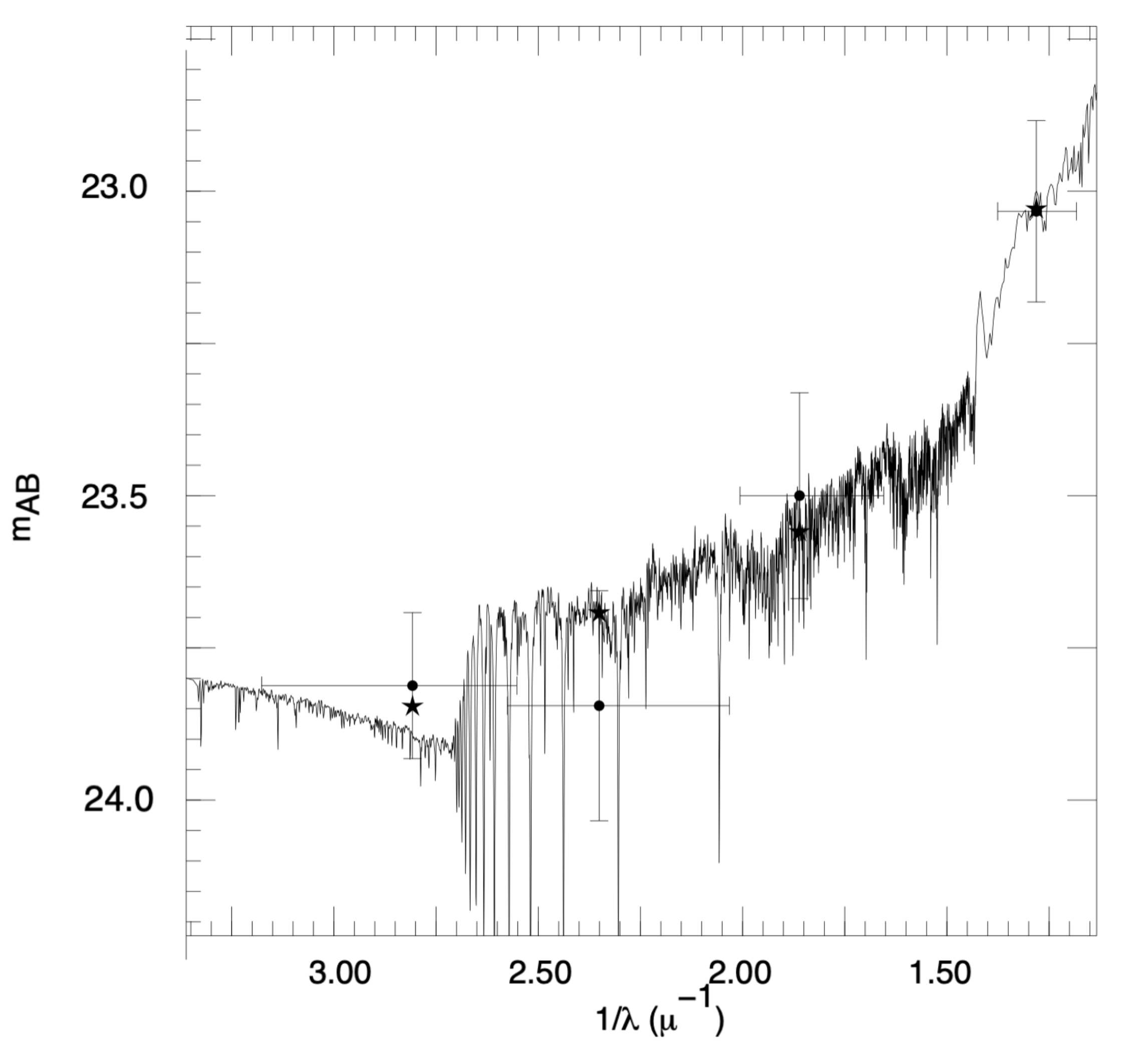}{0.5\textwidth}{(a)}
          \fig{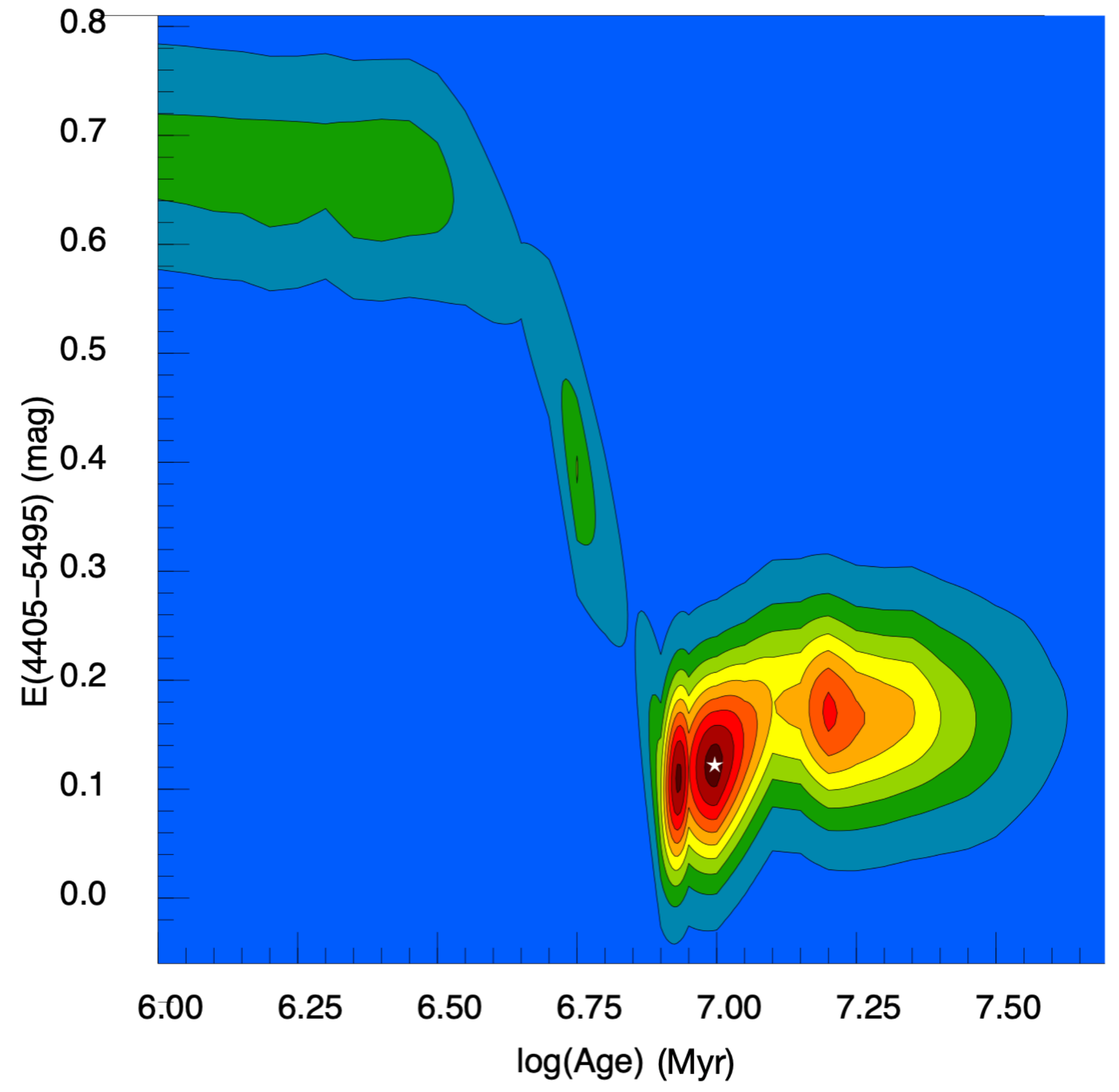}{0.5\textwidth}{(b)}}
\caption{{\sc chorizos} fits to host cluster of SN~2013ai. (a) STARBURST99 model fitted to the photometry of the probable host cluster of SN~2013ai. (b) Chi-squared plane of the fit.}
\label{fig:fit}
\end{figure*}

To estimate an age for Source A, the {\sc chorizos} SED-fitting package was used \citep{Maiz2004}. A grid of Starburst99 models was fit \citep{Leitherer1999} to the measured WFPC2 photometry of the source, constraining the reddening-law parameter R$_{5495}$ to be 3.1, and the metallicity to the solar value, while allowing the cluster age and extinction to vary. The results of the fitting procedure are shown in Figure \ref{fig:fit}. The best fitting model (shown in the left panel) has an age of 7.9 Myr and a relatively low extinction of E(4405-5495)=0.12 mag. Comparing to the STARS models\footnote{https://people.ast.cam.ac.uk/~stars/} \citep{Eggleton2011}, such a cluster age would imply a progenitor mass of $\sim$21 \msun; however, we note that there are also reasonable fits for 11.2 Myr, giving a lower progenitor mass of $\sim$17 \msun\ and higher extinction. The low mass best fit of $\sim17$ \msun\ is at the high end of SN~II progenitor masses \citep{Smartt2009,Sukhbold2020}.

The extinction towards the cluster inferred by the fit is considerably lower than that seen towards SN~2013ai itself. There is a tail of solutions in the $\chi^2$ plane which accommodates a reddening of  E(4405-5495)=0.7 mag, which is consistent with that seen towards the SN. However, these solutions require the cluster to be younger than $\sim$3 Myr, which would in turn lead to a progenitor which is extremely massive, and probably inconsistent with the maximum mass of a star that can maintain its H envelope to the point of core-collapse.

Further analysis was done using the Bayesian fitting technique of \citet{Maund2018}, which addresses cluster mass, distances, and filter transformations. Both the WFPC2 photometry and WFPC2 photometry converted to Johnson-Cousins photometry were included as seperate runs. Given the luminosity of the cluster, the mass is constrained to lie in the range $\log(M) = 10^{4.5} - 10^{5.5}M_{\odot}$.  At such masses, individual, luminous stars may have a disproportionate influence on the colour of the cluster that might cause deviations from the expected colour-sequence predicted by STARBURST99 (for much more massive clusters of $10^{6}M_{\odot}$).

The progenitor of SN~2013ai could be a single, isolated red supergiant as found for other Type II SNe. The relatively bright cluster nearby makes the identification of such a progenitor difficult, as the flux in the wings of the cluster may mask that of a fainter progenitor close by. The most convincing demonstration of such a progenitor would rely on image subtraction, where a flux deficit in new HST images taken after SN~2013ai has faded would indicate the absence of progenitor flux which was present in pre-explosion images. As demonstrated by \cite{Maund2014} and \cite{Folatelli2015}, difference imaging using late-time data can substantially improve the precision and sensitivity of limits placed on SN progenitors. While such a test will have to wait for new observational data, it is still possible to estimate the sensitivity of pre-explosion data to a single red supergiant progenitor. From photometry of point sources close to the SN position, we estimate a $5\sigma$ limiting magnitude of F814W $\gtrsim$ 24.5 mag for the WFPC2 images. At the distance of NGC 2207, this would imply an absolute magnitude for the progenitor of F814W$>$-9 mag, which unfortunately only allows us to constrain the luminosity of the progenitor to log L/\lsun $<$5.5.  For single star models, this luminosity corresponds to a progenitor zero age main sequence mass of $<$24 \msun\ \citep{Eldridge2004}.

\section{Modeling}
\label{sect:model}
\subsection{Modeling Background}
Determining the properties of CCSNe poses a challenge due to the wide range of parameters, e.g. progenitor mass, mass loss pre-explosion, explosion mechanism, mixing, and possible interaction, which affect the observational features.
However, using a combination of empirical measurements and modeling we attempt to reconstruct these properties for SN~2013ai and motivate the use of a SN~1993J inspired model.

The chemical and density structures of the outer layers of the progenitor are revealed by the spectral evolution.
In general, at early epochs, the spectral features are formed at or above the electron scattering photosphere in the H-rich layers. 
During this time, the photosphere can be traced with measurements of weak lines that form near the photosphere, e.g. Fe~{\sc ii}. 
The inner edge of the H-rich layers is determined by the point in time when the Doppler shifts of the Balmer lines becomes constant, around 40 days past explosion in SN~2013ai, whereas Fe~{\sc ii} still receeds. 
At the photosphere, the ejecta density profile of a SN~II can be approximated by a power law, much like that of a stellar atmosphere, $\rho=r^{-n}$, with $n$ the density slope and $r$ the radius at a point in the ejecta \citep{Hoeflich1990}. 
Before the recombination phase of H, the differential Doppler shift between Balmer series H-lines can be used to reconstruct the density structure by integration of the density and slope \citep[see][for applications to SNe~II]{Hoeflich1988,Hoeflich1990}.
The result of applying this technique to SN~2013ai gives a density slope, n, ranging from $\sim$12-25.
A typical SN~II has a density slope of $\sim$10 at early times \citep[][]{Hoeflich1990}, significantly flatter than that of SN~2013ai.
The steep density profile, H line profiles, and Doppler shifts are more similar to SN~1993J than the other SNe compared to (see Figures \ref{fig:optCompareComb} and \ref{fig:NIRcompare}) with both SNe having photospheric velocities of $\sim$12,500 \kms\ at the time of the transition between the H- and He-rich layers, suggesting a similar specific energy in the H-rich layers.

Around a week after maximum, Fe~{\sc ii} is used to trace the photosphere.
This is around the same time that a sharp drop in the Fe~{\sc ii} $\lambda5169$ velocity is observed, as seen in Figure \ref{fig:Gcompare}. 
The drop in velocity is likely caused by the photosphere receding quickly through the He layers into the carbon/oxygen (C/O) core. He~{\sc i} has a high ionization potential and requires high energy nonthermal photons to be observed \citep{Graham1988}. 
We see little evidence of He~{\sc i} in the optical which supports the notion that the jump in Fe~{\sc ii} velocity is caused by the photosphere quickly passing through the He layer. In the NIR, He~{\sc i}  $\lambda1.083\,\mu m$ is seen, however, this He feature is the basis for many other He transitions. Thus, even an object with little He should show NIR He~{\sc i} $\lambda1.083\,\mu m$ given its high population.

The light curves are powered by the energy deposition due to the propagating shock wave, the radioactive decay of $^{56}Ni\rightarrow$$^{56}Co\rightarrow$$^{56}Fe$, and the recombination energy. 
The rise time is dominated by the diffusion time scales in the optically thick inner layers, and the light curve tail is governed by radioactive decay and, possibly, interaction with the CSM and ISM. 
The rise time of SN 2013ai is typical of a SN~IIb and in particular similar to that of the prototypical SN~IIb SN~1993J, which, together with the spectroscopic similarities highlighted above, motivate us to use SN1993J as a starting point to model SN 2013ai.
SN~1993J was extensively modeled by \citet{Hoeflich1993} and was found to have a layered He-C/O core of $\sim10$ \msun\ with an outer H-rich layer mass of
$\sim$ 3 \msun\ originating from a progenitor with a main sequence mass ($M_{MS}$) of $\sim$ 25 \msun.
Note that the final H, He, and C/O core masses depend on the phase of stellar evolution during which mass-loss occurs. 

\subsection{Modeling Results}
For the progenitor evolution of SN~2013ai, we use the stellar evolution code
{\sl MESA} \citep{Paxton2011}. 
To trigger the explosion, the explosion energy ($E_{exp}$) is put in the inner region as a thermal bomb which causes the ejection of the envelope leaving behind a neutron star or a black hole. 
For computational efficiency and to cover a sufficiently wide range of parameters, we first used the code developed by \citet{Bersten2011} to model SN~2013ai.
To verify and further constrain these models, we then modeled SN~2013ai with the non-LTE hydrodynamical-radiation code HYDRA \citep{Hoeflich2003,Hoeflich2009}.
A wide range of $M_{MS}$, $E_{exp}$, and mixing, including pure C/O cores, were evaluated. We will omit the ``failed" attempts. 

\begin{figure}
    \centering
    \includegraphics[width=\columnwidth]{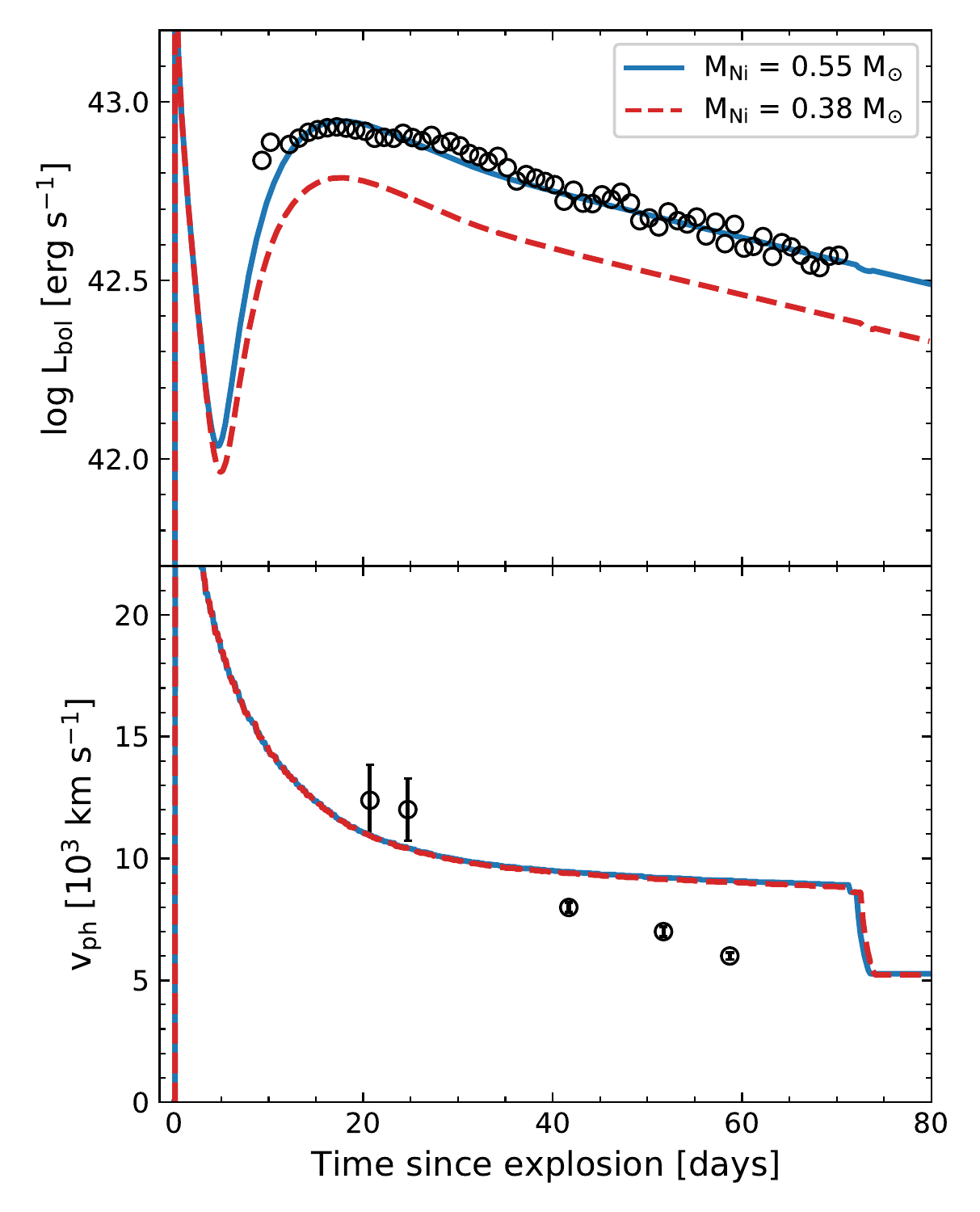}
    \caption{Our preferred light curves and velocities of SN~2013ai using the procedure and codes of \citet{Bersten2011}. The observed photosphere velocties are from the Fe~{\sc ii} $\lambda$5169 absorption feature. The bolometric model shows good agreement with observations. 
    The free parameters were thus constrained to give an explosion energy of 3.0~foe, a $^{56}$Ni mass of 0.55~\msun, and a progenitor mass of 20~\msun. Also shown is a lower $^{56}$Ni mass model to demonstrate that the light curve shape is robust to changes in $^{56}$Ni mass. Note that the early maximum in the model light curves are from shock breakout.}
    \label{fig:grayModels}
\end{figure}

The comparison between a frequency independent model from the code of \citet{Bersten2011} with the photometry is shown in Figure \ref{fig:grayModels} using depth dependent mean opacities based on SN~1993J and originating from a 20 \msun\ star.
The resulting models have 0.55 \msun\ of $^{56}$Ni. Such a high $^{56}$Ni mass is problematic for the explosion of a massive star both from observations and theory (see \citealt{Thielemann2018} for a review). 
The bolometric light curve, in principle, is a sensitive measure of the physical parameters \citep[e.g.][]{Suntzeff1992,Bersten2011}.
Despite the advantage of the bolometric light curve, three problem zones may be identified: 1) the reconstruction of the bolometric light curve, 2) distance uncertainties, and 3) asymmetries. Moreover, using the bolometric light curve reduces the information for finding model parameters. Due to these uncertainties, we use monochromatic light curves based on detailed non-LTE models.

\begin{figure}
    \includegraphics[width=\columnwidth]{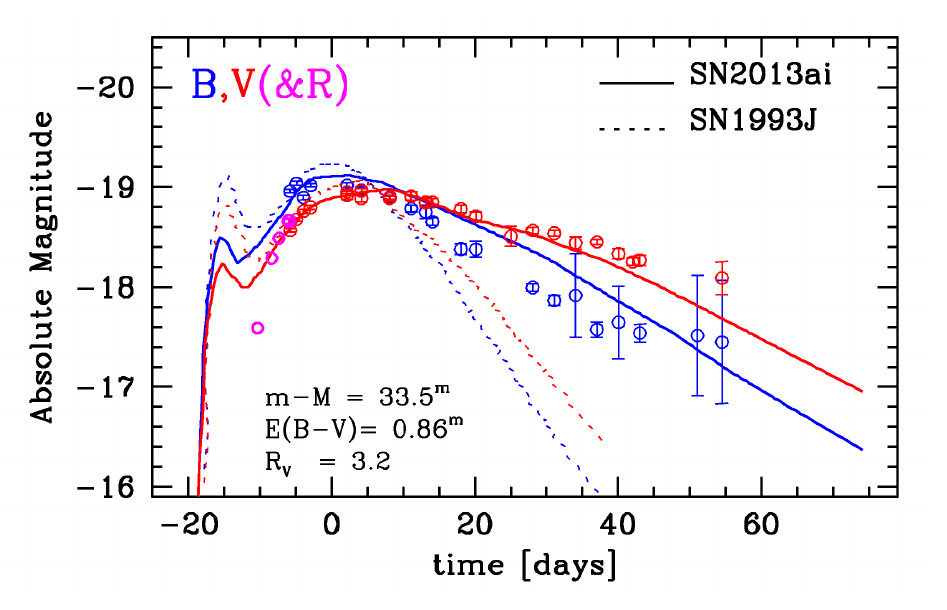}
    \caption{Spherical model calculations for the $B$ and $V$ light curves with the same progenitor model used for SN~1993J, made using non-LTE code with full-radiation transport \citep{Hoeflich2003,Hoeflich2009}. Early $R$-band data is included as a proxy for $V$.
    Free parameters were $E_{exp}=2.0$~foe, the mass of the H-shell $M_{H}=0.2$~\msun,  $M_{Ni}=0.3$~\msun, and Rayleigh-Tailor mixing of the stellar core that gave $v_{RT} = 2500$~\kms after the explosion. 
    Note that the early peak in these models is not due to a UV flash, but arises from energy stored in the H-rich shell at early times, similar to what was seen in observations of SN~1993J.}
    \label{fig:NLTECompare}
\end{figure}

\begin{figure}
    \includegraphics[width=\columnwidth]{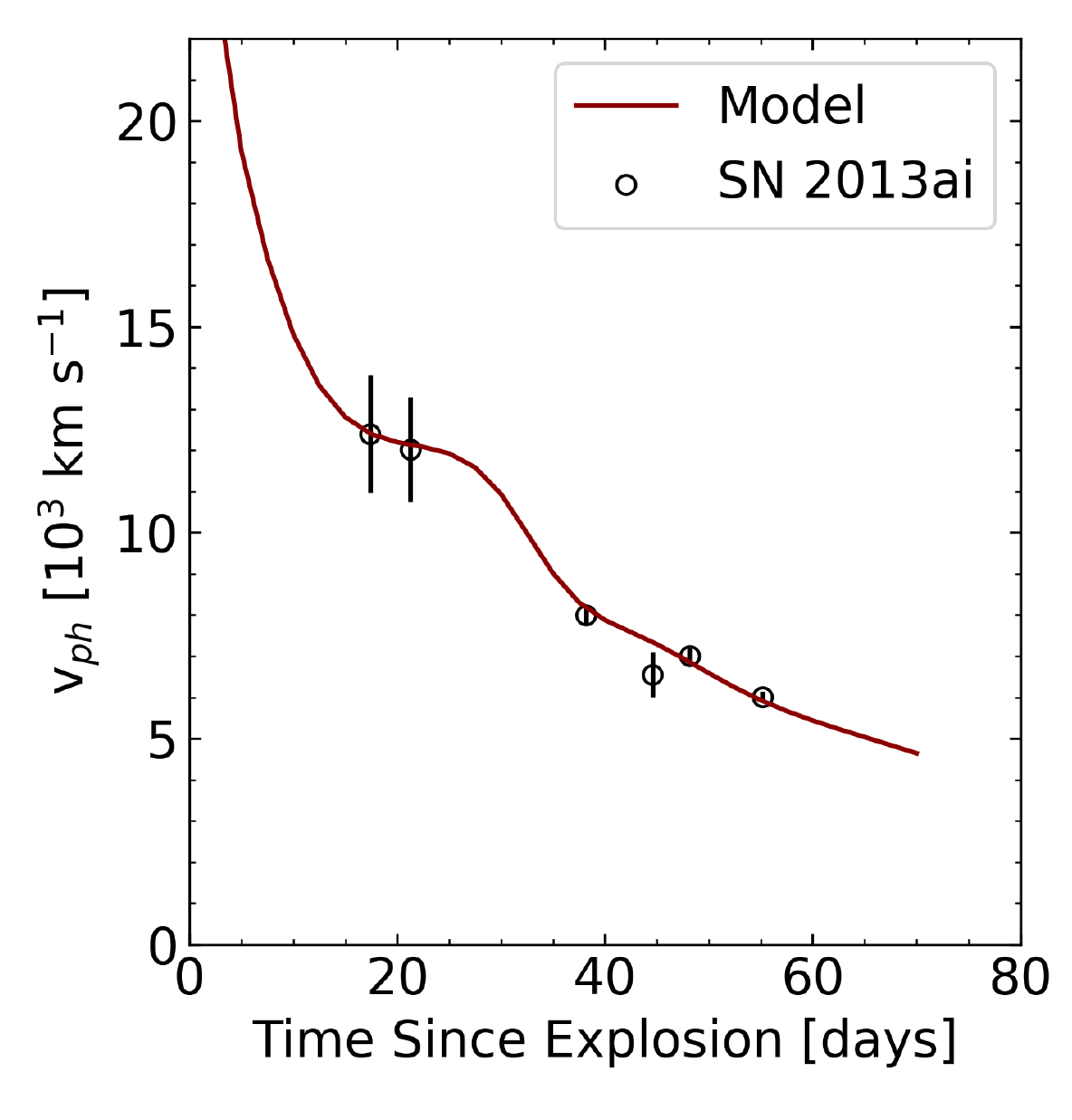}
    \caption{Photospheric radius as a function of time for the non-LTE explosion model shown in Figures \ref{fig:NLTECompare} and \ref{fig:asphCompare} using the Rossland-averaged opacities in V.
    The observed velocties are from the Fe~{\sc ii} $\lambda$5169 absorption feature. The rapid drop seen in the velocity marks the transition from the H-rich to He-rich layers.}
    \label{fig:HYDRA_vel}
\end{figure}

In Figures \ref{fig:NLTECompare} and \ref{fig:HYDRA_vel}, we show the results of HYDRA starting with the progenitor of SN~1993J but modifying the explosion energy, the mass of the H-rich layers, the explosion energy, and the $^{56}$Ni mass. The H-rich layer mass was adjusted to 0.2~\msun, and the $^{56}$Ni mass to 0.3~\msun. These models are based on the results from the previously mentioned models, hereafter referred to as ``LC models". 
The results between best fit models of our simulations, presented as (LC, HYDRA), give consistent results within the uncertainties: $M_{MS} = (20,30)$ \msun, $E_{exp} = (3.0,2.5)$ foe\footnote{1 foe = $10^{51}$ ergs}, and $M_{Ni} = (0.55,0.3)$ \msun.

\begin{figure}
    \includegraphics[width=\columnwidth]{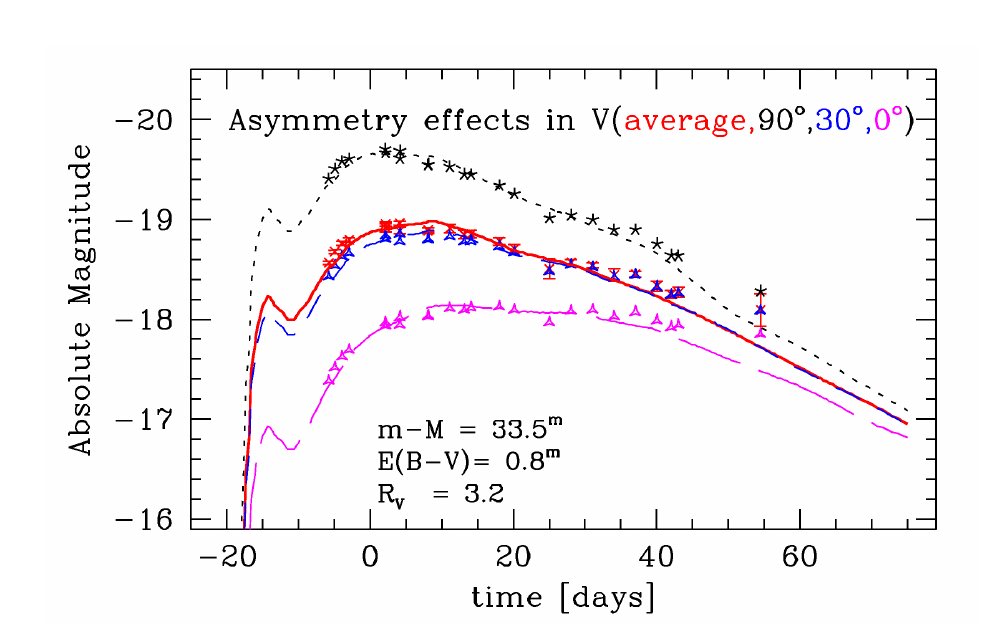}
    \caption{Potential influence of  asymmetry on the $V$-band light curve of SN~2013ai if seen from the pole, $30^o$, and the equator (top to bottom) and
    for both the observations (dots) and model (lines) (see Fig. \ref{fig:NLTECompare}). 
    Both the rise time and magnitude are highly dependent on viewing angle at early times, however, the light curve becomes more robust to asymmetry at later times.
    For the asymmetry, we use the redistribution function of jet-driven explosion models with a difference of a factor of two in the directional dependent 
    explosion which has been used for SN~1998bw, an extreme example \citep{Hoeflich1999}.}
    \label{fig:asphCompare}
\end{figure}

Given the model best fits, the effect of uncertainties on the $M_{Ni}$, $E_{exp}$, C/O core mass, and distance modulus must be discussed. Linear polarizations give evidence for a symmetry axis and bipolar explosions with  $P \approx 1\%$ indicating axis ratios; a measure of how non-spherical an object is, of 1/2-1/4 \citep{Wang1998,Maund2007a,Maund2007b,Stevance2016,Reilly2017}.
To estimate the uncertainty for the light curves, we use SN~1998bw/GRB980425 as a proxy for the explosion of a massive stripped envelope SN \citep{Hoeflich1999}.
Detailed analysis of polarization spectra resulted in an axis ratio of 0.7 for SN~1998bw seen from an angle of $30^o$ \citep{Hoeflich1995} but with significant uncertainties as discussed in \citet{Stevance2020}. 
The possible effect of asymmetry on the $V$-band light curve of SN 2013ai is shown in Figure \ref{fig:asphCompare}. Asymmetry may change the shape of the light curve by varying the maximum brightness up to $\pm1$ mag, placing SN~2013ai near superluminous SNe (see Figure \ref{fig:asphCompare}), and can shift the rise time by about 10 days.

Even for the same model, the light curve rises more slowly and shows a lower maximum brightness that directly translates into uncertainties in $E_{exp}$, and the C/O core mass. 
The error for the asymmetry and the shape may be regarded as an upper limit because even for purely jet-driven explosions of CCSNe, an extended outer envelope tends to become more spherical with time \citep{Khokhlov1999,Hoeflich2004,Couch2009}.
The colors are found to be less sensitive as they are a measure of the physical conditions at the photosphere, namely the flux density rather than the total flux. 
In either case, the tail of the light curve is less effected by polarization because the luminosity becomes isotropic within $\sim 0.2$ mag 1-2 months after the explosion, which translates directly into an error for $M_{Ni}$. 
For the $^{56}$Ni mass, both 0.3 and 0.55 \msun\ fit the data; however, as previously mentioned $0.3-0.55$~\msun\ of $^{56}$Ni is higher than expected for a massive star explosion.

Though beyond the physics included in our simulations, note that the theoretical $V$-band light curve at 250 days is fainter by $\sim$ 1 mag than observed. 
This may indicate a significant  contribution by interaction between the SN-ejecta and the ISM or a pre-explosion stellar wind. 
This theory would also be consistent with the multi-tiered H$\alpha$ emission profile seen in the nebular spectrum of SN~2013ai, which indicates interaction.

\section{Discussion}
\label{sect:discussion}

Given the similar early time light curve timing, expansion velocities, and density structure to SESNe, we suggest that the core of SN~2013ai is closer in physical properties to a SESNe than a normal SN~II.
This would give SN~2013ai a dense $\sim$ 10~\msun\ C/O core and little hydrogen ($0.2$~\msun\ from models) in its outer layers, consistent with the quick decline post-maximum. 
In addition, the explosion energies of SNe~2013ai and 1993J are comparable, 2.5 foe and 2.0 foe, respectively. 
In this scenario, the progenitor of SN~2013ai would have most of its hydrogen envelope stripped before explosion, much like a SN~IIb. 


This brings us to the differences between SN~2013ai and SESNe. The first difference is seen in the shape of the early time, $\sim$2 weeks pre-maximum, light curve.
SN~1993J, for example, shows an initial maximum and subsequent decline in the light curve lasting about one week (see Figure \ref{fig:13aiVS}) that is powered by the energy deposited by the shock in the very outer layers. This structure of the SN~2013ai theoretical LC is very similar to that observed in SN~1993J but this phase has not been observed in SN~2013ai.
In the SN~2013ai HYDRA models, the early peak is much less pronounced because the lower H-shell mass and the longer diffusion times with limited mixing, as predicted by \cite{Bersten2012}.
Secondly, the $^{56}$Ni mass is significantly larger in SN~2013ai than in SN~1993J, which could affect the lack of strong He~{\sc i} seen in the optical.
However, in SN~2013ai, the much stronger He~{\sc i} line in the  NIR $\lambda1.083\,\mu m$ is present and shows the presence of a He layer. 
To boost the optical He~{\sc i} lines in SN~1993J, models with $^{56}$Ni mixed into the outer layers of the ejecta were needed.
This is also consistent with the rapid drop in velocity seen at about 1 week after maximum because the continuum opacity is low.
Finally, SN~2013ai shows a  much slower post-maximum decline rate over 30 days ($\sim$0.5~mag) compared to the rapid decline in SN1993J, ($\sim$2~mag). 
The difference can be attributed to strong $^{56}$Ni mixing in SN~1993J that results in a steeply increasing $\gamma $-ray escape, whereas SN~2013ai has almost full trapping.
While the models based on \citet{Bersten2011} show general agreement with the observations, the strong mixing needed is not consistent with little He~{\sc i} seen in the optical.

From observations, SN~2013ai seems to be rather unusual but, overall, it fits well within the picture of an energetic explosion of a massive star which lost most (but not all) of its hydrogen envelope, which puts it at the edge of becoming a SN~Ib or SN~Ic, physically. However, using the observational classification scheme of SNe, SN~2013ai lies between a SN~II and a SN~IIb as its spectra are that of a SN~II and its light curve more similar to SESNe.
Explosions of massive stars may appear either as SNe~II, IIb, Ib, or Ic depending on the details of the mass loss and the amount of mixing.
The open question remains whether the mass loss in SN~2013ai is line driven or the result of a binary evolution. However, given its similarities to a SN~IIb we cannot rule out that mass loss pre-explosion occurred in a binary system for SN~2013ai.
As such, the current classification scheme may mask diversity and similarity as SN~2013ai has the spectra of a SN~II but after analysis of spectra and light curves SN~2013ai lies between SNe~II and SNe~IIb.

\section{Conclusions}
\label{sect:conclusions}

SN~2013ai is a rare SN that exhibits a long light curve rise, sustained high expansion velocities, and a light curve decline similar to a SN~IIL.
Only one SN~II was found to have comparable light curves and spectra, ASASSN-14kg. However, the lack of ASASSN-14kg data prevents a more detailed comparison between the two SNe.
Some similarities were seen in light curve shape and expansion velocities when comparing SN~2013ai with SNe~IIb.
The data presented provide a link between different classes of SNe. In the current classification scheme, SN~2013ai lies between SNe~II and SESNe with possible signs of interaction seen in the late-time H$\alpha$ profile and early X-ray observations.
The mass loss during the stellar evolution may be a decisive factor.
However, similar data sets for a large number of SNe~II are needed to better describe details of the explosion physics and to search for the diversity and address the question of the origin of this mass loss.
We demonstrated the importance of the combination of optical and NIR spectra with light curves but, in lack of spectropolarimetry, we had to rely on similarity arguments and were unable to constrain some errors present in the early light curve data due to asymmetry. 

\acknowledgements

We thank the staff at the different observatories for performing the observations. Some of the data presented in this paper were obtained from the Mikulski Archive for Space Telescopes (MAST). STScI is operated by the Association of Universities for Research in Astronomy, Inc., under NASA contract NAS5-26555. Based on observations made with the 6.5 meter \textit{Magellan} Telescopes at Las Campanas Observatory, Chile. Based on observations made with the Gran Telescopio Canarias (GTC), installed at the Spanish Observatorio del Roque de los Muchachos of the Instituto de Astrofísica de Canarias, in the island of La Palma. Based on observations collected at the European Organisation for Astronomical Research in the Southern Hemisphere, Chile as part of PESSTO, (the Public ESO Spectroscopic Survey for Transient Objects Survey) ESO program ID 188.D-3003. Based on observations made with the the Very Large Telescope (VLT) and Nasmyth Adaptive Optics System (NAOS) Near-Infrared Imager and Spectrograph (CONICA) program ID 090.D-0329. 

The work of the CSP-II has been generously supported by the National Science Foundation under grants AST-1008343, AST-1613426, AST-1613455, and AST1613472. The CSP-II was also supported in part by the Danish Agency for Science and Technology and Innovation through a Sapere Aude Level 2 grant. MF is supported by a Royal Society - Science Foundation Ireland University Reseach Fellowship. P.H. acknowledges support by grants of the NSF AST-1008962 and Nasa's ATP-1909476. L.G. was funded by the European Union's Horizon 2020 research and innovation programme under the Marie Sk\l{}odowska-Curie grant agreement No. 839090. This work has been partially supported by the Spanish grant PGC2018-095317-B-C21 within the European Funds for Regional Development (FEDER). T.W.C. acknowledges the EU Funding under Marie Sk\l{}odowska-Curie grant agreement No 842471. M.S. is supported by generous grants from Villum FONDEN (13261, 28021) and by a project grant (8021-00170B) awarded by the Independent Research Fund Denmark.

\facilities{Magellan:Baade, 
Swope, 
MAST, 
GTC, 
VLT, 
HST, 
TAROT, 
NTT, 
CTIO:1.3m}

\software{PyWiFeS \citep{Childress2013}, 
IRAF \citep{Tody1986}, 
SNooPy \citep{Burns2011}, 
Astrodrizzle \citep{Hack2012}, 
chorizos \citep{Maiz2004}),
MESA \citep{Maiz2004}, 
HYDRA \citep{Hoeflich2003,Hoeflich2009}, 
STARS \citep{Eggleton2011}.}

\end{document}